\documentstyle[12pt]{article}
\def\figcap{\section*{Figure Captions\markboth
        {FIGURECAPTIONS}{FIGURECAPTIONS}}\list
        {Figure \arabic{enumi}:\hfill}{\settowidth\labelwidth{Figure
999:}
        \leftmargin\labelwidth
        \advance\leftmargin\labelsep\usecounter{enumi}}}
 \relax
\newskip\humongous \humongous=0pt plus 1000pt minus 1000pt
\def\caja{\mathsurround=0pt} \def\eqalign#1{\,\vcenter{\openup1\jot
\caja   \ialign{\strut \hfil$\displaystyle{##}$&$
\displaystyle{{}##}$\hfil\crcr#1\crcr}}\,} \newif\ifdtup
\def\panorama{\global\dtuptrue \openup1\jot \caja
\everycr{\noalign{\ifdtup \global\dtupfalse     \vskip-\lineskiplimit
\vskip\normallineskiplimit      \else \penalty\interdisplaylinepenalty \fi}}}
\def\eqalignno#1{\panorama \tabskip=\humongous
\halignto\displaywidth{\hfil$\displaystyle{##}$
\tabskip=0pt&$\displaystyle{{}##}$\hfil
\tabskip=\humongous&\llap{$##$}\tabskip=0pt     \crcr#1\crcr}}
\def\frac#1#2{ {{#1} \over {#2} }}
\def\sfrac#1#2{\mbox{\small $\frac{#1}{#2}$}}
\def\half{\mbox{\small $\frac{1}{2}$}}
\def\tird{\mbox{\small $\frac{1}{3}$}}
\def\ltap{\raisebox{-.4ex}{\rlap{$\sim$}} \raisebox{.4ex}{$<$}}
\def\gtap{\raisebox{-.4ex}{\rlap{$\sim$}} \raisebox{.4ex}{$>$}}
\def\VEV#1{\left\langle #1\right\rangle}

\def\ie{\hbox{\rm i.e. }}
\def\etal{\hbox{\rm et al.}}
\def\g{\gamma}
\def\as{\alpha_S}
\def\asb{\bar \alpha_S}
\def\bas{\bar \alpha_S}
\def\blank#1{{\hbox {\hskip #1 true cm}}}   
\def\abs#1{\left| \: #1 \: \right|}%
\def\bom#1{\mbox{$\bf #1$}}
\def\beq{\begin{equation}}
\def\eeq{\end{equation}}
\def\re#1{(\ref{#1})}
\def\rg {\right\} }
\def\lg {\left\{ }
\def\o{\omega}
\def\D{\Delta}
\def\Ft{{\tilde F}}
\def\tF{{\cal F}}
\def\cF{{\cal F}}
\def\cA{{\cal A}}
\def\Tb{{\bom T}}
\def\Jb{{\bom J}}
\def\qb{{\bom q}}
\def\kb{{\bom k}}
\def\pb{{\bom p}}
\def\bp{{\bar p}}
\def\bq{{\bar q}}
\def\xt0{$x \to 0$}
\def\x1{$x \to 1$}
\def\Q_s{\mu}

\def\np#1#2#3{Nucl.\ Phys.\ B#1 (19#3) #2}
\def\pl#1#2#3{Phys.\ Lett.\ #1B (19#3) #2}
\def\pr#1#2#3{Phys.\ Rev.\ D #1 (19#3) #2}
\def\prep#1#2#3{Phys.\ Rep.\ #1 (19#3) #2}

\def\rmp#1#2#3{Rev.\ Mod.\ Phys.\ #1 (19#3) #2}

\def\zp#1#2#3{Zeit.\ Phys.\ C#1 (19#3) #2}
\begin{document}
\begin{titlepage}
\begin{flushright}
     IFUM 486-FT\\
     November, 1994
\end{flushright}
\par \vskip 10mm
\begin{center}
{\Large \bf
QCD coherence in the structure function\\
and associated distributions at small $x$
}\footnote{Research supported in part by the Italian MURST
and the EC contract CHRX-CT93-0357
}
\end{center}
\par \vskip 1cm
\begin{center}
        \par \vskip 2mm \noindent
        {\bf Giuseppe Marchesini}\\
        \par \vskip 2mm \noindent
        Dipartimento di Fisica, Universit\`{a} di Milano,\\
        INFN, Sezione di Milano, Italy
\end{center}
\par \vskip 2mm
\begin{center} {\large \bf Abstract} \end{center}
\begin{quote}

We recall the origin of angular ordering of soft parton  emission
in the region of small $x$ and show that this coherent structure
can be detected in associated distributions.
For structure functions at small $x$ and at fixed transverse
momentum the angular ordering is masked because of the complete
inclusive cancellations of collinear singularities for \xt0.
Therefore, in this case the dependence on the hard scale is lost
and the angular ordered region becomes equivalent to multi-Regge
region in which all transverse momenta are of the same order.
In this limit one derives the BFKL equation.
In general such a complete cancellation does not hold for the associated
distributions at small $x$.
The calculation, which requires an analysis without any collinear
approximation, is done by extending to small $x$ the soft gluon
factorization techniques largely uses in the region of large $x$.
Since one finds angular ordering in the both regions of small and
large $x$, one can formulate a unified evolution equation for the
structure function, a unified coherent branching and jet algorithm
which allows the calculation of associated distributions in all
$x$ regions.
Such a unified formulation valid for all $x$ is presented and compared
with usual treatments.
In particular we show that the associated distributions at small $x$
are sensitive to coherence.
By replacing angular ordering with the multi-Regge region one neglects
large singular contributions in the associated distributions.
\end{quote}
\end{titlepage}

\section { Introduction}

Soft gluons emission in perturbative QCD takes place into angular
ordered regions \cite{IR}-\cite{CFMO}.
Since this coherent emission factorizes, one can extend the branching
process, which sums collinear singularities to leading (and next-to-leading)
order, to include angular ordering.
This coherent branching process allows one to compute parton distributions
either by numerical methods via Monte Carlo simulations
or by analytical methods via ``jet calculus'' algorithms
\cite{IR,jetC}.
Experiments are well compatible with coherence and for a recent
review see Ref.~\cite{W}.
An important case in which soft gluons are involved is deep inelastic
scattering (DIS) in the region of large $x$
with $x=Q^2/2pq$ and $Q^2=-q^2$ (see fig.~1).
Coherence in the structure function $F(x,Q)$ for \x1 can be taken
into account in the Altarelli, Parisi, Gribov, Lipatov, Dokshitzer
(AP) equation \cite{DGLAP}
by using as evolution variable the angle of emitted partons.
Coherent branching at large $x$ reproduces \cite{CMW} the leading
and next-to-leading
contribution of the anomalous dimension $\g_\o(\as)$
\beq\label{sf}
\g_\o(\as) = Q^2 \frac{\partial \ln\, F_\o(Q)}{\partial Q^2} \,,
\;\;\;\;\;\;
F_\o(Q)=\int_0^1 dx \,x^\o\, F(x,Q) \,,
\eeq
in the limit $\o \to \infty $ which corresponds to the region of
large $x$.

The small $x$ region in DIS appears quite different from the one of
large $x$ (or finite $x$ in general). For \xt0 one focus the
attention on the soft
exchanged gluons $k_1, \cdots, k_r$ rather than on the emitted ones
$p_1, \cdots, p_r$.
The basis for the analysis of the small $x$ structure function is the
Balitskii, Fadin, Kuraev and Lipatov (BFKL) equation \cite{BFKL}
which was found almost two decades ago in the field of soft physics to
discuss the rise of the total cross section.
The dominant phase space region considered is the so called
''multi-Regge'' region in which subenergies are large and all emitted
transverse momenta are of the same order of magnitude.
In general, if the ratio of transverse momenta vanishes, one has
collinear singularities. However for fully inclusive distributions all
these collinear singularities completely cancel, so that the regions of
very different transverse momenta are not particularly important.

Later it was realized \cite{BFKL,GLR} that this equation could be extended
to describe the structure function for \xt0.
This equation involves the unintegrated structure function
$\cF(x,\kb^2)$ related to the structure function by
\beq\label{sfk}
F_\o(Q)=\int d^2 k \cF_\o(\kb^2)\, \Theta(Q-k) \,,
\;\;\;\;\;\;
k \equiv \abs{\kb}\,,
\eeq
where $\kb$ is the total transverse momentum of emitted gluons,
$\kb_r=\sum \bom p_{ti}$.
For $\o \to 0$ one probes the small $x$ region and in this limit
the $\kb$-structure function satisfies the BFKL equation
\beq\label{BFKL}
\cF_\o(\kb)=\delta^2(\kb)+
\frac{\bas}{\o}\int \frac{d^2p_t}{\pi p_t^2}\,
\left[\cF_\o(\kb+\bom p_t)-\Theta(k-p_t)\,\cF(\kb)\right]
\,,
\;\;\;\;\;
\bas \equiv \frac{C_A\as}{\pi}
\,,
\eeq
with $C_A=3$. The corresponding $\o \to 0 $ anomalous dimension is
given by an implicit equation
and has an expansion in $\as^k/\o^k$. The first six terms are
\beq\label{adim}
\g_\o(\as)=\frac{\bas}{\o}+
2\zeta(3)\; \frac{\bas^4}{\o^4}+
2\zeta(5)\; \frac{\bas^6}{\o^6}+
\cdots
\,,
\eeq
with $\zeta(3) = 1.202 \dots$ and $\zeta(5) = 1.037 \dots$ the
Riemann functions. For the expansion up to $14$ terms see \cite{CFM}.
As well known $\g_\o(\as)$ has a singularity at
$\o=\o^*=4\ln 2\,\bas $
which implies that the structure function rises asymptotically
at small $x$.
The two terms in the kernel of \re{BFKL} have a simple
interpretation.
The first is positive and is the contribution for the emission of
a gluon of transverse momentum $\bom p_t$. Thus, before the emission,
the $\kb$-structure function has momentum $\kb+\bom p_t$.
The second negative contribution in the kernel corresponds to the
virtual correction in which the virtual transverse momenta $\bom p_t$
is bounded by the total emitted transverse momentum $\kb$.
There are two related features which make the BFKL equation apparently very
different from the AP equation.
The first is that there is no any ordering in the real emission
and virtual momenta.
The second is that the collinear singularities for $p_t \to 0$ present
in the real and virtual contributions cancel completely.
This last feature can be further understood if we consider the
exclusive
representation of the $\kb$-structure function in eq.~\re{BFKL}.
To this end we need to introduce a cutoff $\mu$ both for the real
and the virtual contribution and the BFKL equation becomes equivalent
to the exclusive representation
(see the kinematical diagram in Fig.~2)
\beq\label{BFKLe}
\eqalign{
&
\cF_\o(\kb)=\delta^2(\kb) + \sum_{r=1}\int_{\mu^2}^{Q^2} \prod_1^r
\frac{d^2p_{ti}}{\pi p_{ti}^2}
\,dz_i\,z^\o\,\frac{\bas}{z_i}\,\Delta_R(z_i,k_i)
\;\delta^2(\kb-\kb_r)\,,
\;\;\;\;\;\;\;
\kb_\ell=\bom p_{t \ell} +\kb_{\ell-1}
\cr&
\Delta_R(z_i,k_i)=
\exp\left\{-\bas\int_{z_i}^1 \frac{dz}{z}\,
\int_{\mu^2}^{k_{i}^2} \frac{dp_t^2}{p_t^2}
\right\}
=
\exp\left\{
-\bas\ln\, \frac{1}{z_i}
\,\ln\,\frac{k^2_{i}}{\mu^2}\,\Theta(k_{ti}-\mu)
\right\}
\,,
\;\;\;\;\;
k_i=\abs{\kb_i}
\,.
}
\eeq
The function $\Delta_R(z_i,k_i)$ sums all virtual contributions and
corresponds to the so called gluon ``Regge trajectory''.
In this exclusive representation we have included, together with the
collinear cutoff $\mu$, also the upper limit for the emitted transverse
momenta given by the hard scale $Q$ (see \re{sfk}).
Due to the complete cancellation of collinear singularities
at the fully inclusive level and to the scale invariance of the
measure $d^2p_t/p_t^2$, the $\mu$ and $Q$ dependence in the $\kb$-structure
function \re{BFKLe} enters only as higher twist
corrections, \ie as powers of $\mu^2/Q^2$ and for
$\mu^2/Q^2 \to 0$ the representation \re{BFKLe} is
equivalent to the BFKL equation \re{BFKL}.
Since $Q$ enters only to higher twists one has that for \xt0
the hard scale is lost to the present level of accuracy.
In DIS for \xt0 the hard scale $Q$ is recovered only through the
phase space boundary in the $\kb$ dependence of the coefficient
function via the high energy (or $\kb$) factorization (see
\cite{HEF}).
On the other hand to the present accuracy the hard scale for
the argument of $\as$ can not be identified.
Since only the coefficients of $\as^k/\o^k$ are known, a scale change
between $Q$ and $Q'$
$$
\frac{\as(Q')}{\o}=
\frac{\as(Q)}{\o}
+\frac{\as^2(Q)}{\o}\;2b\; \ln(\frac{Q}{Q'})\; +\cdots
\,,
\;\;\;\;\;\;\; b=(11C_A-2n_f)/12\pi
$$
can not be discriminated.
The control of the hard scale can be obtained by the calculation
of subleading corrections $\as\, \as^k/\o^k$ in the anomalous
dimension.
The first steps in this direction are reported in Ref.~\cite{FL}.

{}From the above discussion it seems that the dynamical features
of the two DIS regions of small and large $x$ is quite different.
However it was shown that the two regions have the same structure
of coherence.
The calculation of the structure function for \xt0 was done
\cite{C}-\cite{CFMO} in the framework of hard processes thus considering
not only the soft singularities for \xt0 but also the collinear
singularities when the ration of two emitted transverse momenta
vanishes. This region is beyond the multi-Regge region.
The result was the discovery of coherence for \xt0. One finds that
for \xt0 initial state soft gluons are emitted into the angular
ordered region as in the case of large $x$. In this way the
description of DIS is unique for all region of $x$.

Notice the accuracy required for this calculation.
Since in fully inclusive distributions all collinear singularities
cancel for \xt0, one needs to perform a calculation without making
any collinear approximation. In the light-cone expansion
infinite leading twists operators contribute and one has to make
a calculation to all loop accuracy.
Actually, by using the soft gluon factorization techniques for \xt0,
it was shown \cite{CFMO} that it is possible to attain such
an accuracy by appropriate use of soft gluon coherence and
of cancellation of soft singularities between real-virtual contributions.
One can use soft gluon techniques also for \xt0 because this region is
dominated by soft exchanged gluons $k_1,\cdots,k_r$ and this implies
that, a part for the first fast gluon $p_1$, all remaining emitted
gluons are soft.

Many data are now available from Hera in the small $x$ region and
then the analysis of coherence at small $x$ can be attempted.
To help this analysis, in this paper we present a systematic account
of coherence in all $x$ regions.
In particular, by using the coherent branching process discussed
above we show how computes the associated distributions
via a generalization of jet calculus \cite{IR,jetC}.
This allows one to study quantitatively the r\^ole of coherence for
\xt0 and to single out the differences between the present formulation
with coherence and the exclusive representation in \re{BFKLe}
or the usual AP evolution equation naively extended to small $x$.

The result of this analysis in the case of small $x$
\cite{C}-\cite{CFMO}
can be summarized in terms of very
simple extension of the exclusive representation of the BFKL
$\kb$-structure function in \re{BFKLe}.
It involves the following two modifications:
\par \noindent
i) the emission takes place in the angular ordered region. When all
exchanged gluons are soft angular ordering can be expressed by including
a factor $\Theta(p_{ti}-z_{i-1}p_{t\,i-1})$ in the phase space for each
emission in \re{BFKLe}.
The collinear cutoff is needed only for the first emission with $i=1$,
\ie $p_{ti} > z_0p_{t0}=\mu$
(this cutoff is not needed for the $\kb$-structure function);
\par \noindent
ii) also the virtual corrections involve angular ordering. The form
factors $\Delta_R(z_i,\kb_i)$ in \re{BFKLe} are modified by
substituting the cutoff $p_{t}>\mu$ by the angular ordered
constraint $p_{t}>z p_{ti}$.

On the basis of this branching structure one deduces \cite{C,CFM}
an evolution equation for the $\kb$-structure function
at small $x$ which involves as evolution an angular variable.
{}From this equation one observes that there is a complete cancellation
of the collinear singularities. In this case angular ordering becomes
equivalent to the multi-Regge region \re{BFKLe}.
On the contrary we show that in the associated distributions,
such as the associated multiplicity, the collinear singularities
do not cancel. In  this case one can not use the multi-Regge region
in \re{BFKLe} but must take into account angular ordering and
the above modifications.

The fact that both for small and large $x$ parton emission and virtual
corrections factorizes and have the same structure of coherence, \ie
angular ordering, allows one to deduce a unified branching process for
all $x$ regions.
This has been used to construct a Monte Carlo program \cite{MW}
including coherence in the small $x$ region which has been used to
study the differences with the one loop formulation
and to compute numerically DIS and heavy flavour production processes.
Moreover one can write \cite{ERICE} a unified evolution equation for
the $\kb$-structure function which extends the AP equation to include the
coherence at small $x$ \cite{C,CFM}.
We show that this equation, which \xt0 reduces to the BFKL equation, can
be used as
the bases for the jet calculus to compute the associated
distributions.

In Sec.~2 we recall the results of Refs.~\cite{C}-\cite{CFMO} of the
DIS analysis for \xt0 together with \x1. We describe the accuracy of
the approximations in the various region of $x$.
In Sect.~4 we discuss the branching process with coherence for all $x$
and in Sect.~5 we recall the unified evolution equation
\cite{ERICE} for the $\kb$-structure function valid in all $x$
regions.
In Se.~6 we discuss the small $x$ region. At the fully inclusive level
we show that from the evolution equation with coherence one deduces
the BFKL equation valid in the multi-Regge region.
At the semi-inclusive level (associated distributions) we show that
the present formulation with coherence can not be approximated
by using only the multi-Regge region as in the BFKL equation.
In Sect.~7 we describe the extension of jet calculus
to include the coherence at small $x$ and compare with a
formulation in the multi-Regge region.
Finally Sect.~8 contains some conclusion.

\section {Derivation of recurrence relation  and solution}

We consider deep inelastic scattering represented at parton
level in Fig.~1
where $\, q \,$ is the hard colour singlet probe and $\, p' \,$ represents
the recoiling system of partons, such as, for instance, a pair of heavy
quark-antiquark.
For simplicity we assume that the incoming
$\, p \,$ and the outgoing partons $\, p_1 \ldots p_r \,$ are gluons.
We take $p=(E,0,0,E)$ and introduce a light-like vector
$\bp=(E,0,0-E)$ in the opposite $z-$direction.
For the emission momenta we use the Sudakov parametrization
\begin{equation}\label{mom}
p_i = y_i p +\bar y_i \bar p +p_{ti} \, ,
\;\;\;\;\;\;\;\;
\xi_i \equiv \frac {\bar y_i}{y_i} =\; \frac {p^2_{ti}}{s\, y_i^2}
\, ,
\;\;\;\;\;\;\;\;
s=2p \bar p \, ,
\end{equation}
where $\ln \xi_i$ is the (pseudo-)rapidity.
For the exchanged momentum one has $k= x p + {\bar x} \bp +k_t$.
We assume the frame $q=\bp-xp$ with $x=Q^2/2pq$ and consider
$Q^2=-q^2$ large. In this limit we are interested in the region
where all $\bar y_i$ are small.

The cross section of this process factorizes into the elementary
hard scattering distribution and the structure function.
For $x \to 0$ this factorization can be generalized \cite{HEF}
to include higher order corrections by introducing the structure
function at fixed $\kb $.
To leading twist, in the gauge $\, \eta = \bp \equiv q+xp \sim p'$,
the amplitude of Fig.~1 can be factorized \cite{CFM,CFMO}
as in Fig.~3
into the elementary vertex ${\bom V}^\mu(q,k)$ and the multi-gluon
emission subamplitude ${\bom m}^{(r)}_{\mu}(p,k;p_1, \ldots p_r)$
which is a matrix in the colour indices of gluons
$p, p_1, \cdots p_r$ and $k$.
In this gauge, for \xt0, the elementary vertex $\, \bom V^\mu$,
is proportional to $\bar p^\mu$ and one defines the multi-gluon
subamplitudes
$ {\bom M}_{r}(p_1,\ldots,p_r) =
({\pb^\mu}/{2p \bp}) \;{\bom m}^{(r)}_\mu(p,k;p_1,\ldots,p_r)$.
One introduces the structure function at fixed total transverse
momentum
\begin{equation} \label{Ft}
\eqalign{
&\kb^2\,\Ft(x, \kb) =  \frac 1 x \sum_r \frac {1}{r!} \int
\prod_{i=1}^r (dp_i) \abs{M_{r}  (p_1 \ldots p_r)}^2
\delta (1-x-Y_r) \delta^2 (\kb - \kb_r) \; , \cr
& (dp)   =  \frac{d^3 p}{(2\pi)^3 2 \omega}\: ;
\blank{.6}
Y_r = \sum_1^r y_i \: ;
\blank{.6}
\kb_{r} = - \sum_{i=1}^r {\pb}_{it}
\,,
}
\end{equation}
with $\, \abs{M_{r} (p_1 \ldots p_r) }^2 $
the spin and colour average of the multi-gluon squared subamplitude.
The structure function is obtained by integrating over $\kb$
\begin{equation} \label{F}
F(x,Q) = \int d^2\kb_t \;\tF(x,\kb,Q)\,,
\;\;\;\;\;\;\;\;
\tF(x,\kb,Q) \equiv \Theta(Q - k) \Ft(x, \kb) \,,
\;\;\;\;\;\;\;\;\;
k\equiv \abs{\kb}\,.
\end{equation}
By integrating over $y_i$ in \re{Ft} one finds infrared (IR)
divergences which are cancelled by virtual contributions.
These cancellations are most easily seen in old fashion perturbation
theory in which virtual gluons are on-shell.
To order $\, \as^n $, the distribution $\, \abs{M_{r} }^2 \,$
is given by integrals over $\ell=n-r$ on-shell virtual momenta
$v_1 \cdots v_\ell$.
We introduce the on-shell momenta $\{ q_1,\ldots,q_n \}$ which
includes both the $r$ real $\{ p_1 \ldots,p_r \}$ and
the $\ell=n-r$ on-shell virtual
$\{ v_1 \ldots v_{\ell} \}$ momenta
and we use the Sudakov parametrization
$q_i=y_ip+\bar y_i \bar p + q_{ti}$ with
$\bar y_i=q^2_{ti}/y_is$.
One has to consider contributions with various ordering in the
$y_i$ variables of the on-shell real and virtual momenta
and for a given contribution we denote by
$ f_{\,r,\ell}\;(p_1 \ldots p_r; v_1 \ldots v_{\ell})$
the corresponding integrand in which the spin and colour sum is
already performed.
We then introduce the index $ R_n \,$ which identifies the real
$\{p_i\}$ or virtual $\{v_i\}$ momenta in the set $\{q_i\}$ and
denote by $ f_{R_n} (q_1 \ldots q_n)$ the corresponding integrand.
We then have then the following representation of the structure
function.
\begin{equation}
\label{calF}
\eqalign{
&
x\,\tF(x, \kb,Q)  =
\sum_{n=1}^\infty g_S^{2n}
\int \prod_{i=1}^n (dq_i)
\; \Theta(Q-q_{it})
\; \Theta^Y_{1\ldots n}
\; \sum_{R_n} f_{R_n} (q_1 \ldots q_n)
\cr&
\times
\delta \left( 1-x-Y_{R_n} \right)\,
\delta^2 \left( \kb - \kb_{R_n} \right) \; ,
\;\;\;\;\;\;\;\;\;\;\;\;
\Theta_{1 \ldots n}^Y \equiv \theta(y_1-y_2)\cdots
\theta(y_{n-1}-y_n) \, ,
}
\end{equation}
where $Y_{R_n}=Y_r \,$  and $\, \kb_{R_n}=\kb_{r} \,$ are
the contributions from the real gluons and are defined in \re{Ft}.
Obviously the integrand $\, f_{R_n} (q_1 \ldots q_n)$ depends
on the specific index $R_n$.
The bound $Q > q_{it}$ is coming from $Q^2 > \kb^2 $ in \re{F} and
the ordering of the $y_i$ variables is coming both from the
statistical factor $1/r!$ for real momenta and from all possible
ordering of the virtual on-shell momenta.

We need now to compute $f_{R_n}(q_1, \ldots q_n)$.
We will do this in the strongly $y$-ordered region
\begin{equation} \label{yo}
y_n \ll \cdots \ll y_1 \,,
\end{equation}
in which we can apply the soft gluon techniques and
factorize the contribution of the softest gluon, either real or
virtual.
This region gives the leading contribution both for $x \to 0$ and
for $x \to 1$.
The fact that for $x \to 1$ the strongly ordered phase space gives
the dominant contribution is well known \cite{IR}.
The fact that the ordering \re{yo} gives  also the dominant
contribution for $x \to  0$ is due to the infrared regularities of
the inclusive distributions.
Since in \re{calF} the various singular contributions for $y_i \to 0 $
cancel, the integration regions $y_i < x $ give a contribution to the
structure function of order $x$ which can be neglected for \xt0.
On the other hand in the remaining regions $y_i > x$
the real-virtual cancellations are inhibited by the presence of
$\delta(1-x-Y_{R_n})$ in the integrand.
Therefore $x$ plays the role of an infrared cutoff and logarithmic
contributions in $x$ are generated by the lower bound in the $y_i$
integrations.

In the remaining part of this section we shall recall \cite{C,CFM,CFMO}
that by factorizing the softest (real or virtual) gluon $q_n$ one deduces
the following recurrence relation connecting $f_{R_n}$ and $f_{R_{n-1}}$
\beq\label{rr}
\eqalign{
&
\sum_{R_n}
\, f_{R_n} (q_1 \ldots q_n)
\, g^2_S\, (dq_n)
\, \delta ( 1-x-Y_{R_n}) \delta^2( \kb - \kb_{R_n})
\, \simeq
\, \sum_{R_{n-1}} f_{R_{n-1}} (q_1 \ldots q_{n-1})
\, \bas
\, \frac {dy_n}{y_n}\frac{d^2\qb_n}{\pi \qb_n^2}
\cr&\times
\biggl\{
\delta ( 1-x-Y_{R_{n-1}}-y_n)
\, \delta^2( \kb - \kb_{R_{n-1}} + \qb_n) \;
\cr&
-\delta ( 1-x-Y_{R_{n-1}} )
\, \delta^2 ( \kb - \kb_{R_{n-1}})
\, \left[ 1- \Theta(y_n-x)\Theta(q_n-k) \right]\,
\biggr\}
\,.
}
\eeq
As we shall recall, this recurrence relation is valid in the infrared
limit
(\ie $y_n \ll y_i$)  for both \xt0 (without collinear approximation)
and \x1 (in the collinear approximation).
The first contribution in the curly bracket corresponds to the case
in which the softest gluon is emitted,
while the second corresponds to the case in which $q_n$ is a virtual
gluon.
Consequently, the momentum of the softest gluon $q_n$ enters in the
the $\delta-$function only for the first contribution.
This recurrence equation gives rise to an integral equation
for the $k-$structure function.
One needs to fix the total transverse momentum $\kb$ because
this variable enters in the coefficient of the $\delta-$function of
the last term of \re{rr}.

\subsection{Recurrence relation for $f_R$}

We summarize here how to obtain from the soft gluon factorization
the recurrence relation \re{rr} and discuss the relative accuracy.
First we recall the case of large $x$ and then we discuss the case of
$x$ both small and large.

\par\noindent
{\it Large $x$}.
As well known \cite{IR}, the factorization of the softest
(real or virtual) gluon $q_n$ in this region gives the
colour matrix factor
\beq \label{eik}
\eqalign{
\;
-\left[
{\Jb}_{R_{n-1}}(q_n)
\right]^2\,
\biggl\{
&
\delta (1-x-Y_{R_{n-1}}-y_n)\,\delta^2(\kb-\kb_{R_{n-1}}+\qb_n)
\cr&
-
\delta (1-x-Y_{R_{n-1}}    )\,\delta^2(\kb-\kb_{R_{n-1}}      )
\biggr\}
\,.
}
\eeq
This result is valid in the soft approximation ($y_n \ll y_i$)
and no collinear approximations are needed.
In the first contribution $q_n$ is the real emitted gluon thus it
contributes to the $\delta$-functions in \re{calF}.
In the second contribution the gluon $q_n$ is virtual and does not
contribute to the $\delta$-functions.
The soft emission factor is the squared of the eikonal current
\beq \label{J}
{\Jb}_{R_{n-1}} (q_n)=
-\frac{p}{pq_n}\Tb_{p}
+\sum_{i \in R_{n-1}}  \frac{q_i}{q_i q_n}\Tb_{q_i}
+ \frac{p'}{p'q_n}\Tb_{p'} \: ,
\;\;\;\;\;\;\;
\Tb_{p'}= \Tb_p - \sum_{i \in R_{n-1}}  \Tb_{p_i} \,,
\eeq
with $\, \Tb_{q_i} \,$ and  $\, \Tb_{p} \,$ the colour matrices
of external gluons, and $\Tb_{p'}$ the colour matrix of the
recoiling system $p'$.
The emission factor is universal since it depends only on the fastest
charges. therefore it does not depends on particular Feynman graph
contributions. Moreover the factor is the same for $q_n$ real and
virtual.

The current $\Jb$ is a matrix in the space of
$\{p,q_1,\cdots,q_{n-1},p'\}$ colour indices and in the use of \re{J}
the difficulty is the colour algebra.
Since the integrand $f_{R_n}$ is given by the average over the
colour charges, the factorized expression in \re{eik} gives a
recurrence relation for $f_{R_n}$ only if the colour factor in
\re{eik} is approximately proportional to the unit matrix.
In this case of large $x$ this is true only in the collinear limit in
which the emitted gluons $q_i$ are taken parallel to the incoming one,
\ie $q_i\simeq y_i\,p$.
By using colour conservation in \re{J} and $\Tb_{p}^2=C_A=N_C$,
in this limit one has
\beq\label{coll}
-\left[{\Jb}_{R_{n-1}}(q_n)\right]^2
\simeq
-C_A\,\left( \frac{p}{pq_n}-\frac{p'}{p'q_n}\right)^2
\simeq \frac{4C_A}{\bom q^2_{t}}
\,.
\eeq
For large $x$ the coherent emission is obtained in this approximation
to the leading collinear accuracy. Therefore the analysis in
this case is to the leading IR and collinear accuracy.

\par\noindent
{\it Small and large $x$}.
This analysis can be extended to include the region of small $x$.
Here the analysis seems much more difficult since one can not make
any collinear approximation.
Moreover soft gluon factorization  techniques are
typically used in the case in which the softest gluon $q_n$ is softer
than all other external $q_i$ and internal lines.
In the $x \to 0$ case instead we have that
the exchanged gluon $k_n$ could be softer than $q_n$.
Actually the soft gluon technique has been generalized to the \xt0
case in Ref.~\cite{C,CFM} and one finds that the contribution of
the gluon $q_n$ can be factorized in $f_{R_n}$ without the need of
any collinear approximation \cite{CFMO}. There are three factorized
contributions:

\noindent
i) the softest gluon $q_n$ is a real emitted one. One can
generalize
the result \re{eik} also to the small $x$ case and factorize the
colour matrix $-\left[{\Jb}_{R_{n-1}}(q_n)\right]^2$.
It is surprising that also for \xt0 the leading IR contribution for
the real emission is given by the eikonal current involving only the
external partons.
Actually, as shown in \cite{CFMO}, for \xt0 the contribution
of the emission of $q_n$ from the softest exchanged line $k_n$ is
important.
This contribution compensate exactly the rescaling
$\kb_n^2/(\kb_n-q_n)^2$ for the exchanged propagator
and, for \xt0, the emission is given by the squared eikonal current
\re{eik} without any collinear approximation;

\noindent
ii) the softest $q_n$ is a virtual on-shell gluon not connected
to the exchanged gluon $k$. In this case we have again the
squared eikonal current factor in \re{eik};

\noindent
iii) the softest $q_n$ is a virtual on-shell gluon connected to the
exchanged gluon $k$. This factor, which is singular only for
$x \ll y_n$, is positive and proportional to the unit colour matrix.
We have
\beq
\label{ne}
\frac{4C_A}{\qb_n^2}\;
\frac{(\kb+\qb_n)\qb_n}{(\kb+\qb_n)^2} \;
\Rightarrow
\frac{4C_A}{\qb_n^2} \, \Theta(\qb^2_n-\kb^2)
\eeq
where the last expression is obtained upon azimuthal integration.

It is an important consequence of coherence in the limit \xt0
that if one takes together the first two contributions the colour
algebra becomes trivial as in \re{coll} but without any collinear
approximation.
This can be shown by observing first that in the strongly ordered
region \re{yo} and for \xt0 one has $y_1 \to 1$. Thus in this limit
the $\delta$ functions in the $y$ variables are given by
$\delta(1-y_1)$ both in the real and in the virtual contribution.
Therefore, apart for this common $\delta(1-y_1)$ function, the sum of
the real and the first virtual contribution are given by
\beq\label{noncoll}
\eqalign{
\left[{\Jb}_{R_{n-1}}(q_n)\right]^2\,
&
\left\{
\delta^2(\kb-\kb_{R_{n-1}}+\qb_n)
-
\delta^2(\kb-\kb_{R_{n-1}})
\right\}
\cr
\simeq
\frac {4C_A}{\qb^2_n} \,
&
\left\{ \delta^2(\kb-\kb_{R_{n-1}}+\qb_n)
-
\delta^2(\kb-\kb_{R_{n-1}})
\right\}
\,,
}
\eeq
as in the collinear limit. However this expression is valid
in the limit $y_n \ll y_i$ without requiring any collinear
approximation.
To show this observe that for $y_n \ll y_i$ the difference of
the two delta functions gives a non-vanishing contribution only in the
collinear region with $\theta_i \sim y_n/y_i \theta_n$. Thus for
$y_n/y_i \to 0$ we can use the approximate expression \re{coll} for
the squared eikonal current.
Therefore it is the coherence of the soft radiation for \xt0 which allows
us to compute the eikonal current in the collinear limit.
For finite $x$ one can not neglect the differences in the
arguments of the $\delta$ function in $y_i$ for the real and
the virtual contributions and, in this case, the approximation
\re{coll} is valid only if one requires the collinear limit.

In conclusion from the factorization results in \re{ne} and \re{noncoll}
we obtain in the soft limit $y_n \ll y_i$ the recurrence relation
in \re{rr} which is valid both for \xt0 (without collinear approximation)
and for \x1 (in the collinear approximation)
The three $\delta$-function contributions in the curly bracket of
\re{rr} have the
following origin: the first corresponds to the real emission;
the second and third correspond to the eikonal and non-eikonal
virtual correction respectively. The new feature of the last
contribution, which is relevant only for \xt0, is that it involves
the total transverse momentum. This implies that the kernel and
then the solution of the recurrence equation must involve $\kb$.
This is in agreement with the fact that for small $x$ one introduces
\cite{HEF} the structure function at fixed $\kb$.

\subsection{General solution of the recurrence relation}
{}From the recurrence relation \re{rr} one obtains the multi-gluon
distributions $\abs {M_r}^2 $ in \re{Ft}.
The virtual corrections exponentiate and can be summed to give
(see appendix A)
\beq\label{M2}
\frac {1}{\kb^2}\prod_1^r(dp_i)
\;\abs{M_r(p_1,\ldots,p_r}^2
\;\Theta^Y_{1\ldots r}
\;\simeq\;
\bas^r\,
\,\prod_1^r \frac {dy_i}{y_i}\frac {d^2p_{it}}{\pi p^2_{it}}
\;\, V_r(p_1,\ldots,p_r)
\; \Theta^Y_{1\ldots r}
\,,
\eeq
where the virtual corrections are given by
\beq \label{V}
V_r(p_1p_2 \ldots p_r)=
\exp{\lg
-\bas\int^1 \frac{dy}{y} \int^{Q^2}\frac{dq^2}{q^2}
\rg}
\;\; \prod_{i=1}^r \;
\exp{\lg
\asb
\int_{x_i}^{x_{i-1}}
\frac{dy}{y} \;
\int_{\kb^2_i}^{Q^2}
\frac{dq^2}{q^2}\;\;  \rg}
\,,
\eeq
with
$$
{\kb}_{i}=\kb_{i-1}-\pb_{t i} \, ,
\;\;\;\;\;\;\;
x_i=x_{i-1}-y_i \, ,
\;\;\;\;\;\;\;
\kb=\kb_r \,,
\;\;\;\;\;\;\;
x=x_r \;.
$$
The first exponential in $V_r$ is the eikonal form factor which sums
all virtual corrections corresponding to the second
term in the curly bracket of \re{rr}.
Notice that for $y \to 0$ this integral diverges. Similarly all
real integrations diverge for $y_i \to 0$. These IR singularity
cancel in inclusive quantities such as the structure function.
However, if we want to keep separately real and virtual
contributions, one should regularize these IR singularities
by a momentum transverse cutoff.
This will be carefully done later.

The remaining factor in \re{V} is given by the form factors which sum
the contribution from the last term in the curly bracket of
\re{rr}, corresponding to the non-eikonal virtual corrections.
These form factors give singular contributions only in the case
of soft exchange ($x_i \ll x_{i-1}$), \ie for \xt0.

\section{Coherence and angular ordering}
The expression of the squared amplitudes $\abs{ M} ^2$ in \re{M2}
is given in the strongly $y$-ordered region \re{yo}.
In order to obtain an evolution equation one
has to exchange $y$-ordering with some other ordered variable
involving the transverse momentum or angle of emitted gluon.
The identification of this variable allows us to exhibit the
coherence properties of these distributions.
One can show that within the leading IR accuracy one can replace
in \re{M2} the ordering in $y$ with the ordering in
the angles between the incoming gluon and the emitted
ones, \ie the ordering in $\xi$ (see Appendix B). We have
\beq\label{V1}
\prod_1^r \frac {dy_i}{y_i}\frac {d^2p_{it}}{\pi p^2_{it}}
\; V_r(p_1p_2 \ldots p_r)
\;\Theta^Y_{1\ldots r}
\;\; \Rightarrow \;\;
\prod_1^r \frac {dy_i}{y_i}\frac {d^2p_{it}}{\pi p^2_{it}}
\; V_r(p_1p_2 \ldots p_r)
\; \Theta^{\xi}_{r\ldots 1}
\, ,
\eeq
where
$$
\Theta^{\xi}_{r \ldots 1} \equiv
\Theta(\xi_r-\xi_{r-1}) \ldots \Theta(\xi_2-\xi_{1})\,,
$$
so that the gluons $\{p_1 \ldots p_r\}$ are now emitted in the
angular ordered phase space region
\begin{equation}\label{ao}
\xi_1 < \ldots < \xi_r  .
\end{equation}
Since the $y$-variables are not any more ordered we introduce
the usual $z_i$ variables
$$
y_i=(1-z_i)x_{i-1}\,,
\;\;\;\;\;\;
x_i=z_1x_{i-1}\,,
\;\;\;\;\;\;
x=x_r=z_1\cdots z_r\,,
$$
so that the $y$-integration factors in \re{M2} can be written
\beq\label{zdist}
\frac {\asb^r}{x}
\;\prod_1^r \; \frac{dy_i}{y_i}
\; \delta(1-x-\sum_1^r  y_i)=
\;\prod_1^r \; dz_i
\;\; \asb \lg \frac {1} {1-z_i} +\frac {1} {z_i} \rg
\; \delta (x- z_1 \cdots z_r) \,.
\eeq
We recognize the singular parts of the gluon splitting function. For
$z_i \to 1$ we have soft emission and for $z_i \to 0$ soft exchange
which is relevant for \xt0.
The finite contribution to the gluon splitting function
$-2+z_i(1-z_i)$ can be included. They can not be obtained within the
soft gluon techniques.

The IR singularities $z_i \to 1$ present in the real emission
distribution \re{zdist} are cancelled, at the inclusive level, by the
virtual corrections in $V_r$.
If we want to keep separately the real and virtual contributions we
need to regularize these IR singularity
by introducing a cutoff $Q_0$ in the transverse momenta.
As usual we introduce the rescaled
transverse momenta \beq\label{q}
q_i \equiv \frac{p_{ti}}{1-z_i} =x_{i-1} \sqrt{s \xi_i}\,,
\end{equation}
so that the angular ordered region \re{ao} becomes
\beq\label{ao1}
\xi_i > \xi_{i-1}\,,
\;\;\;\;\;\;
\xi_1>\xi_0
\;\;\; \Rightarrow \;\;\;
q_i > z_{i-1}q_{i-1}\,,
\;\;\;\;\;\;
q_1>\Q_s
\,,
\eeq
where we have introduced also the collinear cutoff $\xi_0$ or
$\Q_s=\sqrt{s\,\xi_0}$.
The IR singularities in \re{zdist} for $y_i \to 0$
(\ie $\,z_i \to  1$) are regularized by limiting the emitted
transverse momentum, $p_{it} > Q_0$, giving
\beq\label{Q0}
1-z_i > Q_0/q_i\,.
\eeq
In terms of these variables, the virtual corrections in \re{V} can
be written, to leading IR order, in the form (see Appendix C)
\beq\label{V11}
V_r(p_1p_2 \ldots p_r)
\simeq
\,\Delta_S(Q,\,z_rq_r) \prod_1^r \Delta_S(q_i,\,z_{i-1}q_{i-1})
\;\Delta(z_i,q_i,\kb_{i}) \, ,
\end{equation}
where $\Delta_S$ is the usual gluon Sudakov form factor in the double
logarithmic approximation
\beq\label{sud}
\Delta_S(q_i,z_{i-1}q_{i-1})=
\exp \left\{ -
\int_{{(z_{i-1}q_{i-1})}^2}^{q_i^2 }
\frac{dq^{ 2} }{q^{2} }
\int_{0}^{1 - Q_0/q} {dz}
\frac { \asb } {1-z}  \,
\right\} \,.
\end{equation}
Notice that the $q$ integration region ($q>z_{i-1}q_{i-1}$)
corresponds to the angular ordering constraint \re{ao1}.
Here the regularization of IR singularity $(1-z) > Q_0/q$ is
the same as in the real emission phase space.

The other non-Sudakov form factor is given by
\beq\label{nsud}
\Delta(z_i, q_i, \kb_{i})
= \exp \lg
-\asb \int_{z_i}^1 \frac {dz}{z}
\int^{\kb^2_{i}} \frac {dp^2}{p^2}\Theta(p-zq_i)
\rg \,,
\;\;\;\;\;\;
\kb_i \equiv \sum_{\ell=1}^i \pb_{\ell t}\,.
\end{equation}
The upper limit $\kb_{i}^2$ in the $p^2$ integration comes from the
non-eikonal
form factor in \re{V}, while the lower limit $zq_i$ corresponds
to the angular ordering $\xi > \xi_i\,$
($ p = z\, x_{i-1}\,\sqrt{s\, \xi} >z q_i \, $)
and comes from a part of the eikonal form factor not included
into $\Delta_S$ (see Appendix C).
This form factor is different from the Sudakov form factor in
various respects.
The Sudakov form factor is associated to IR singularities for
the virtual on-shell gluon thus it regularizes the $z \to 1$
singularity in the splitting function of the usual evolution equation.
It is a function of the local variables at the vertex such as
the two angular variables in \re{sud}.
The form factor in \re{nsud} is associated to a singularity
in the soft exchange gluon $z_i \to 0$. It depends on the energy
fraction $z_i$, the angular variable $q_{i}$ and the exchanged
transverse momentum $\kb_i$. It is this $\kb$-dependence which
gives one of the important feature for coherence in the small
$x$ region.
For $x$ finite \re{nsud} gives a subleading correction which
is not present in the usual Sudakov form factor even at subleading
level \cite{IR,CMW}.
This form factor is related to the Regge form factor of the gluon
$\Delta_R(z_i,\kb_i)$ which is found in the BFKL analysis \re{BFKLe}.
Here the lower bound integration over
$p$ is bounded for $z \to 0$ by a collinear cutoff $\Q_s$.
Due to this connection we shell call the non-Sudakov form factor
$\D(z,q,\kb)$ also the hard Regge form factor.

\section{Unified coherent branching}
In conclusion, from \re{M2}, \re{V1}, \re{zdist} and \re{V11} we can write
the $\kb$-structure function in eq.~\re{calF} in the form
(see the kinematical diagram in Fig.~2)
\beq\label{calF1}
\eqalign{
&
\cF(x;\kb,\kb_0;Q,\Q_s)  =
\delta(1-x)\,\delta^2(\kb-\kb_0)
\; \Delta_S(Q,\Q_s)\Theta(Q-\Q_s)
+
\sum_{r=1}^\infty \int_{0}^{Q^2}
\Delta_S(Q,z_rq_r)
\cr&\times
\prod_{i=1}^r \;
\left\{
\frac{d^2q_i}{\pi q_i^2}\, dz_i \,
\bas P(z_i,q_i,k_i)\,
\Delta_S(q_i,z_{i-1}q_{i-1})\,
\Theta(q_i-z_{i-1}q_{i-1})
\right\}
\;\delta(x-x_r)\,\delta^2(\kb-\kb_r)
\,,
}
\eeq
where $x_i=z_ix_{i-1}\;$, $\kb_i=-\qb_i+\kb_{i-1}\;$ and
$k_i=\abs{\kb_i}$.
We have included an incoming transverse momentum $\kb_0$ and the
collinear cutoff $\Q_s=z_0 q_0$ which corresponds to the factorization
scale for collinear singularities.
The splitting function is
\beq\label{br}
P(z_i,q_i,k_i) = \frac{ \Theta(1-z_i-Q_0/q_i) }{1-z_i} +
\frac{\Delta(z_i,q_i,k_{i})}{z_i}
\;.
\end{equation}
We have included the non-Sudakov form factors $\Delta$ only in
the $z_i \to 0$ singular contribution of the gluon splitting
function \re{zdist} since $\Delta$ is regular for finite $z_i$
and gives a non leading correction.
It is easy to include in $P(z,q,k)$ the finite terms $-2+z_i(1-z_i)$
which are relevant in the region of $x$ not large or small.
It is also possible to generalize the equation and the splitting
function to include the quark contributions.
The variable $Q$ has been introduced in \re{F} as the upper bound on
the total $\kb$ (see \re{sfk}). To the present accuracy this is equivalent
to set $Q$ as the upper bound of all emitted transverse momenta.

The main feature of this branching is angular ordering
which is present in the real emission phase space
($q_i > z_{i-1}q_{i-1})$ (see \re{V1}),
in the Sudakov form factor ($q > z_{i}q_{i})$ (see \re{sud}),
and in the non-Sudakov form factor ($p>zq_i$) (see \re{nsud}).
The fact that the non-Sudakov form factor $\D(z_i,q_i,k_i)$, relevant
only at small $x$, depends also on the exchanged transverse momentum
$k_i$ is the reason way we need to introduce the unintegrated
structure function.
The branching distribution in \re{calF1} and \re{br} is the basis for
the new Monte Carlo program \cite{MW} which simultaneously takes
into account coherence for large $x$ (to double logarithms) and
for small $x$ (to all loops).

In the small $x$ region eq.~\re{calF1} sum all leading IR singularities,
\ie powers of $\as^n/x\, \ln^{n-1} x$.
This contributions are obtained when all variable $z$ vanish.
Here the lower boundaries for the transverse momenta
vanishes both in the real emission ($q_{i}> z_{i-1}q_{i-1} \to 0$) and
in the non-Sudakov form factor ($p>zq_i \to 0$).
We have than that the collinear singularities arising from the
vanishing of these lower boundaries generate not only logarithms
in the transverse momenta, which give powers of $\ln Q/\Q_s$,
but also logarithms in the variables $z$, which give further powers
of $\ln x$.
Recall that in the  present formulations all these logarithms of collinear
origin are included without approximations.
As we shall discuss later, in the $\kb$-structure function
all collinear singularities fully cancel so that angular ordering
becomes equivalent to the multi-Regge region.
This is way the expansion of the $\kb$-structure function
contains only powers of $\as^n/x\, \ln^{n-1} x$ and
the BFKL equation is obtained.

\section{Unified evolution equation}

The distribution $\tF(x,\kb,\kb_0;Q,\Q_s)$ in \re{calF1} is given in
terms of the angular ordered phase space and therefore one can
deduce an integral or a differential equation in the angular variable.
See Ref.~\cite{C,CFM} for the case \xt0. Here we consider the extension
to the regions of $x$ both small and large.

For $x$ not small one has that all $z_i$ are finite and, to leading
collinear logarithm, angular ordering in \re{ao} is equivalent to
$q_i$ ordering ($q_i>z_{i-1}q_{i-1} \simeq q_{i-1}$).
The evolution equation is obtained by differentiating \re{calF1} with
respect to $Q$ and one obtains the usual AP evolution equation
with coherence \cite{IR}, \ie the leading $\as \ln(1-x)$ contributions
are correctly summed by taking into account the rescaling factor in the
angular evolution variable $q_i=p_{it}/(1-z_i)$.

For \xt0 one has that $Q$ can not play the r\^ole of an angular variable.
Indeed the last branching phase space is $Q > q_r > z_{r-1}q_{r-1}$ so
that, by differentiating with respect to $Q$, we get $Q=q_r$ while
the upper scale for the angular variable must be $z_rq_r$.
Therefore, in order to obtain an evolution equation also for \xt0 one
needs to introduce an additional variable $\bq$ giving this upper scale
for the last angle of the emission
$$
\xi_r<\bar \xi\,
\;\;
\Rightarrow\,
\;\;
z_r q_r < \bq = x\sqrt{s\bar \xi}
\,.
$$
We then introduce the distribution $\cA(x;\kb,\kb_0;\bq;Q,\Q_s)$
defined by \re{calF1} in which we add the constraints
$\bq>z_n q_n$. We have then
\beq\label{calA}
\eqalign{
&
\cA(x;\kb,\kb_0;\bq;Q,\Q_s) \equiv
\delta(x-1)\delta^2(\kb-\kb_0) \,\Delta_S(\bq,\Q_s)\,\Theta(\bq-\Q_s)
+
\sum_{r} \int_{0}^{Q^2}
\Theta(\bq -z_rq_r)
\,\Delta_S(\bq ,z_rq_r)
\cr&\times
\prod_{i=1}^r
\left\{
\frac{d^2q_i}{\pi q_i^2}dz_i\,
\bas P(z_i,q_i,k_i)
\,\Delta_S(q_i, z_{i-1}q_{i-1})\,
\Theta(q_i- z_{i-1}q_{i-1})\,
\right\}
\,\delta(x-x_r)\,\delta^2(\kb-\kb_r)
\,.
}
\eeq
At $\bq=Q$ we have
$$
\cA(x,\kb,\kb_0;\bq=Q;Q,\Q_s) \;=\; \tF(x,\kb,\kb_0;Q,\Q_s)
\,,
$$
and for $\bq<x\,\Q_s$ the integral vanish thus
\beq\label{bc}
\cA(x,\kb,\kb_0;\bq=x\Q_s;Q,\Q_s)=\delta(1-x)\delta^2(\kb-\kb_0)\,.
\eeq
This distribution satisfies the following integral equation
\beq\label{inteq}
\eqalign{
&
\cA(x,\kb,\kb_0;\bq;Q,\Q_s)=
\delta(x-1) \,\delta^2(\kb-\kb_0)
\,\Delta_S(\bq,\Q_s)\,\Theta(\bq-\Q_s)
\cr&
+
\int_{0}^{Q^2} \; \Delta_S(\bq ,zq) \Theta(\bq -zq)
\;\frac{d^2q}{\pi q^2} \frac{dz}{z}
\;\bas\,P(z,q,k)
\;\cA(\frac x z, \kb',\kb_0;q;Q,\Q_s)\,,
}
\eeq
with $ \kb'=\kb+(1-z)\qb$.
The evolution equation valid for large and small $x$ is obtained by
differentiating with respect to the last angular variable
\beq\label{eveq}
\eqalign{
&
\bq^2 \frac{\partial}{\partial
\bq^2}\cA(x,\kb,\kb_0;\bq;Q,\Q_s)
\cr&
= \int_x^1 \frac {dz}{z}
\,\bas\,
\tilde P(z,\frac{\bq}{z},\kb)
\left [
\left( \frac{1}{1-z} \right)_+
\,+\,\frac{ \Delta (z,\frac{\bq}{z},k) } {z}
\right]
\; \cA\left( \frac x z, \kb',\kb_0;\frac \bq z;Q,\Q_s \right)
\Theta(Q-\frac{\bq}{z})
\,,
}
\eeq
where $ \kb'\equiv \kb+\frac{(1-z)}{z}{\bom \bq}$ and
the azimuthal integration over the direction of ${\bom \bq}$ is
understood.
The $z \to 1$ singularity in the kernel is regularized by
the virtual contribution coming by differentiating the
Sudakov form factors in \re{inteq}.
Notice that the finite $-2 +z(1-z)$ should be added if one considers
regular contribution for $z$ away from the boundaries $z=0$ and $z=1$.
This evolution equation can be generalized to include also the quark
contributions.

It is important to notice that \re{eveq} is actually an evolution
equation in the angular variable $\bq/x=\sqrt{s \bar \xi}$ and not
in $\bq$ itself.
This is seen in the fact that the derivative of the
distribution $\cA(x,\kb,\kb_0;\bq;Q,\Q_s)$
in the l.h.s of \re{eveq} at $x\,$ and $\bq$ is given by the
distribution at $x'=x/z\,$ and $\bq'= \bq/z$.
Moreover the boundary condition \re{bc} is given at the minimum angle
$\sqrt{s\xi_o}=\bq/x=\Q_s$.
Of course one can use instead a boundary condition at fixed $\bq$.
We have than that the solution of \re{eveq} is obtained by increasing
$\bq/x$ up to the values $\bq<Q$.

Consider the two limits \x1 and \xt0.
In the first case we recover easily the usual evolution equation for
the structure function $F(x,Q)$ in \re{F} with coherence included
\cite{IR}.
This is seen by observing that for finite $x$ also $z$ is
finite. Recalling that in this case
we required only the leading collinear accuracy, we can
make the following approximations in \re{eveq}:
i)   take as ordering variable $\bq=Q$;
ii)  neglect the non-Sudakov form factor which it is not
singular for finite $z$;
iii) replace in the integrand $\bq/z $ with $\bq$.
With all these simplification we obtain
$$
Q^2 \frac{\partial}{\partial Q^2}\cF(x,\kb,\kb_0;Q,\Q_s)=
\int_x^1 \frac {dz}{z}
\bas\left[
\left(\frac {1}{1-z}\right)_+ +\frac 1 z
\right]
\; \cF( \frac x z, \kb',\kb_0;Q,\Q_s)
\,,
$$
with $\kb'=\kb+(1-z)Q \bom n$ and the integration over the
two-dimensional versor $\bom n$ is understood.
Notice that by neglecting the non-Sudakov form factor, there is no
$\kb$ dependence in the kernel. Therefore we can integrate this
equation over $\kb$ and obtain the usual AP equation for the structure
function (a part for the finite part in $z$ for the splitting
function).
The $\kb$ dependence of $\cF(x,\kb,\kb_0;Q,\Q_s)$ is known \cite{IR}
and is of a Sudakov form factor type.

We consider now \re{eveq} in the small $x$ region where we can
neglect the $1/(1-z)_+$ term in the kernel. We obtain
(see also \cite{C,CFM})
\beq\label{eveq0}
\bq^2 \frac{\partial}{\partial \bq^2}\cA(x,\kb,\kb_0;\bq;Q,\Q_s)=
\int_x^1 \frac {dz}{z}
\frac {\bas}{z} \, \Delta(z,\frac{\bq}{z},\kb)
\; \cA\left( \frac x z, \kb',\kb_0;\frac \bq z;Q,\Q_s \right)
\Theta(Q-\frac{\bq}{z}) \,,
\eeq
with $\kb'\equiv \kb+\frac{\bom \bq}{z}$
Notice that for $Q$ and $xQ< \bq < Q$ the $z-$integration is over
the range $\bq/Q<z<1$.
Therefore this distribution becomes independent of
$\bq$ as $\bq$ approaches $Q$ and satisfies the BFKL equation.
This is analyzed in the next Section.

\section{Coherence at small $x$}
At small $x$ we can neglect in \re{eveq} both the real
and virtual contributions which are singular for $z \to 1$.
The corresponding approximation in \re{calA} gives
\beq\label{calA0}
\eqalign{
&
\cA(x, \kb,\kb_0;\bar q;Q,\Q_s) =
\delta(x-1) \,\delta^2(\kb-\kb_0) \,\Theta(\bq-\Q_s)
\cr&
+
\sum_{r} \int_{0}^{Q^2}
\,\Theta(\bq -z_rq_r)
\prod_{i=1}^r
\left\{
\frac{d^2q_i}{\pi q_i^2}dz_i
\; \frac{\bas}{z_i}\,\Delta(z_i,q_i,k_i)
\; \Theta(q_i- z_{i-1}q_{i-1})\,
\right\}
\; \delta(x-x_r)\,\delta^2(\kb-\kb_r)
\,,
}
\eeq
with $ x_i=z_ix_{i-1}\,$, $\kb_i=\kb_{i-1}-\qb_i\,$ and
$k_i=\abs{\kb_i}$.

As we shall see, in this fully inclusive distribution all
collinear singularities in the real emission phase space
$q_{i} \to z_{i-1}q_{i-1} \to 0$ are compensated by the collinear
singularities in the non-Sudakov form factors \re{nsud} for
$p \to z q_{i} \to 0$.
Therefore at the inclusive level one can simplify the phase space
by substituting angular ordering with the multi-Regge region in
which all transverse momenta are independent and by regularizing
the exclusive collinear singularities in \re{calA0} by the cutoff
$\Q_s$ in both the real emission and the virtual integrals.
After performing the sum the limit $\Q_s \to 0$ is finite.
This corresponds to make the following approximations in \re{calA0}
\beq\label{BFKLlim}
\eqalign{
&
\Theta(q_i\,-\,z_{i-1}q_{i-1})\,
\Rightarrow\,
\Theta(q_i\,-\,\Q_s)\,
\cr &
\Delta(z_i,q_i,k_i)
\, \Rightarrow \,
\Delta_{R}(z_i,k_i)
\,.
}
\eeq
Within this approximation we obtain the following representation of the
$\kb$-structure function
\beq\label{F00}
\eqalign{
&
\cF_{0}(x,\kb,\kb_0;Q,\Q_s) =
\delta(1-x)\,\delta^2(\kb-\kb_0)\,\Theta(Q-\Q_s)
+
\cr&
\sum_{r=1} \int_{\Q_s^2}^{Q^2}  \prod_1^r
\left\{
\frac{d^2q_i}{\pi q_i^2}
\, dz_i
\, \frac{\bas}{z_i} \Delta_{R}(z_i,k_i)
\right\}
\, \delta(x-x_r)\,\delta^2(\kb-\kb_r)
}
\,,
\eeq
in which the phase space is the multi-Regge region (see \re{BFKLe}).

\subsection{Perturbative expansion}
We check explicitly to four loops that the small $x$ structure functions
in \re{calA0} and \re{F00} are the same for large $Q$.
We will see that the real and virtual contributions have
different type of singularities. In the coherent formulations
\re{calA0} we find stronger singularities for $x \to 0$
than in the approximate form in \re{F00}.
All these stronger singularities do cancel in the
fully inclusive sum ($\kb$-structure function), but they do not
cancel for associated distributions.

Consider the contributions to the structure functions for a fixed number
$r$ of emitted initial state gluons
\beq\label{Fo}
F_\o(Q)
\;=\;
\int_0^1 dx\, x^\o\, F(x,Q)
\;=\;
1+\sum_{r=1}F_\o^{(r)}(Q)
\,.
\eeq
In the formulation with coherence the structure function is obtained
by integrating \re{calA0} over $\kb$ and setting $\bq=Q$. For
simplicity we assume $\kb_0=0$ and we obtain
$$
F_\o(Q,\Q_s) = \Theta(Q-\Q_s)
+
\sum_{r=1} \int_{0}^{Q^2}  \prod_1^r
\frac{d^2q_i}{\pi q_i^2}
dz_i \frac{\bas}{z_i} \,z_i^\o\,
\Delta(z_i,q_i,k_i)\Theta(q_i-z_{i-1}q_{i-1})
\,,
$$
with the collinear cutoff $ \Q_s=z_0q_0$.

In the formulation without coherence in \re{F00} we obtain
$$
F_{0\,\o}(Q,\Q_s) = \Theta(Q-\Q_s)
+
\sum_{r=1} \int_{\Q_s^2}^{Q^2}  \prod_1^r
\frac{d^2q_i}{\pi q_i^2}
dz_i \frac{\bas}{z_i} \,z_i^\o\,
\Delta_R(z_i,k_i)
\,.
$$
For the structure function $F_{0\,\o}(Q)$ in \re{F00} without
coherence we readily obtain the perturbative expansion.
For large $Q$ one obtains to four loops ($ T\equiv \ln(Q/\Q_s)$)
\beq\label{per0}
\eqalign{
&
F^{(1)}_{0,\o}(Q)=
2\frac{\bas}{\o}T
- \sfrac{1}{2}(2\frac{\bas}{\o}T)^2
+\sfrac{1}{3}(2\frac{\bas}{\o}T)^3
-\sfrac{1}{4}(2\frac{\bas}{\o}T)^4+\cdots\,,
\cr&
F^{(2)}_{0,\o}(Q)=
(2\frac{\bas}{\o}T)^2
-\sfrac{7}{6}(2\frac{\bas}{\o}T)^3
+ \sfrac{29}{24}(2\frac{\bas}{\o}T)^4
+8\left(\frac{\bas}{\o}\right)^4\,\zeta(3)T
+\cdots \,,
\cr&
F^{(3)}_{0,\o}(Q)=
(2\frac{\bas}{\o}T)^3
-\sfrac{23}{12}(2\frac{\bas}{\o}T)^4
-4 \left(\frac{\bas}{\o}\right)^4\, \zeta(3)T
+\cdots \,,
\cr&
F^{(4)}_{0,\o}(Q)= (2\frac{\bas}{\o}T)^4  + \cdots
\,.
}
\eeq
The perturbative expansion of these contributions is of the form
$$
F_{0\o}^{(r)}(Q)=\sum_{n=r}^\infty\,\;C^{(r)}_{0}(n;T)\;
\frac{\bas^n}{\o^{n}} \,.
$$
Summing all exclusive emission contribution one finds the perturbative
result \re{adim} of the BFKL anomalous dimension up to four loops.
The first correction to the one loop anomalous dimension
is only at the fourth loop.
The Riemann zeta functions are obtained from integrals of type
\beq\label{Rie}
\int^{Q^2}_{\Q_s^2} \frac {d^2q}{\pi q^2}
\ln^2\left( \frac{(\qb-\qb')^2}{\qb'^2} \right)
\simeq
\int_0^{Q^2}\frac{d^2k}{\pi(\kb-\qb')^2}\ln^2(\frac{k^2}{{q'}^2})
= 4\zeta(3) + \sfrac{1}{3}\ln^3(Q^2/{q'}^2)
+{\cal O}(\frac{\Q_s^2}{Q^2})
\,.
\eeq
We neglected the collinear cutoff $\Q_s$ since the
integral is regular for $ q \to 0$.
Moreover we have taken $Q$ as the upper bound both for the emitted
and the exchanged transverse momentum which is valid for $q' \ll Q$.

In the coherent formulation the perturbative contribution to
four loops are
\beq\label{per}
\eqalign{
&
F^{(1)}_{\o}(Q)=
\frac{2\bas}{\o}T
-\frac{(2\bas)^2}{\o^3}T
+3\frac{(2\bas)^3}{\o^5}T
-15\frac{(2\bas)^4}{\o^7}T +\cdots\,,
\cr&
F^{(2)}_{\o}(Q)=
(2\bas)^2\left\{
\sfrac {1}{2}\frac {T^2}{\o^2} +\frac {T}{\o^3}
\right\}
-
(2\bas)^3\left\{
\frac {T^2}{\o^4} +5\frac {T}{\o^5}
\right\}
+
(2\bas)^4\left\{
\sfrac {7}{2}\frac {T^2}{\o^6} +32\frac {T}{\o^7}
+\sfrac{2}{4}\zeta(3)\frac{T}{\o^4}
\right\}
+\cdots\,,
\cr&
F^{(3)}_{\o}(Q)=
(2\bas)^3\left\{
\sfrac {1}{3!}\frac {T^3}{\o^3} +\frac {T^2}{\o^4}
+2\frac {T}{\o^5}
\right\}
-
(2\bas)^4\left\{
\sfrac {1}{2}\frac {T^3}{\o^5} +6\frac {T^2}{\o^6}
+22\frac {T}{\o^7} +\sfrac{1}{4}\zeta(3)\frac{T}{\o^4}
\right\}
+\cdots\,,
\cr&
F^{(4)}_{\o}(Q)=
(2\bas)^4\left\{
\sfrac {1}{4!}\frac {T^4}{\o^4}
+\sfrac{1}{2} \frac {T^3}{\o^5}
+\sfrac{5}{2} \frac {T^2}{\o^6}
+5            \frac {T  }{\o^7}
\right\}
+\cdots
\,.
}
\eeq
The Riemann zeta functions are obtained from the same integrals in
\re{Rie} with the lower bound $\Q_s$ in general substituted by $zq'$.
The perturbative expansion of $F_\o^{(r)}(Q)$ is of the form
$$
F_\o^{(r)}(Q)=\sum_{n=r}^\infty\,\sum_{m=1}^{n}
\;C^{(r)}(n,m;T) \; \frac{\bas^n}{\o^{2n-m}}
\,.
$$
The singular terms with $m<n$ are related to coherence since
they are coming from angular ordering, \ie the $z$ dependence in the
emission phase space ($q_i > z_{i-1}q_{i-1}$) and in the virtual
integration ($p>zq_i$).
By summing all the exclusive contributions one finds that all these
more singular terms cancel and obtains the same anomalous dimension
dimension \re{adim} to four loop.
This is an example of the complete cancellations of the collinear
singularities in the $\kb$-structure function.

At a less inclusive level, such as for the associated distributions,
the collinear singular terms $({\bas^n}/{\o^{2n-m}})$ with $m<n$
do not cancel any more.
In the approximation \re{BFKLlim} instead, the perturbative coefficients
are given only by powers $(\bas/\o)^n$.
Therefore the approximation of neglecting angular ordering by simply
putting the cutoff $\Q_s$  (see \re{BFKLlim}) is not any more valid for
the associated distributions.

As an example we consider the average number of
gluons emitted in the initial state, \ie we neglect the final state
branching of these gluons. This average is obtained by summing the
contributions \re{per0} and \re{per} with a weight given by $r$.
For the formulation with and without coherence
we find
\beq\label{rungs}
\eqalign{
&
\sum_r\, r\,F^{(r)}_{\o}(Q,\Q_s)\,=
\frac{2\bas}{\o}T +\left(\frac{2\bas}{\o}\right)^2(T^2+\frac{T}{\o}) +
\cdots
\cr&
\sum_r\, r\,F^{(r)}_{0\,\o}(Q,\Q_s)=
\frac{2\bas}{\o}T +\frac{3}{2}\left(\frac{2\bas}{\o}\right)^2\,T^2 + \cdots
}
\,.
\eeq

\subsection{Relation with the BFKL equation}
We want to show now that the structure functions at fixed $\kb$
satisfy the BFLK equation for \xt0.
Consider first the case of no coherence obtained by
approximating angular ordering with the multi-Regge region.
The unintegrated structure function $\cF_{0\o}(\kb,\kb_0;Q,\Q_s)$
is given in \re{F00}.
We show that the dependence on $Q$ and $\Q_s$ is of
higher twist type and, neglecting this dependence, the
unintegrated structure function satisfies
the BFKL equation.
{}From \re{F00} we have
\beq\label{inteq0}
\cF_{0\o}(\kb,\kb_0,Q,\Q_s)=\delta^2(\kb-\kb_0)\Theta(Q-\Q_s)+
\int^{Q^2}_{\Q_s^2} \frac{d^2q}{\pi q^2}dz\,z^\o\,
\frac{\bas}{z}\Delta_R(z,k)
\cF_{0\o}(\kb+\qb,\kb_0;Q,\Q_s)
\,,
\eeq
and one readily deduces the following integral equation
\beq\label{BFKL1}
\eqalign{
\cF_{0\,\o}(\kb,\kb_0;Q,\Q_s)
&
= \frac{\bas}{\o}\int^{Q^2}_{\Q_s^2} \frac{d^2q}{\pi q^2}
\left\{
\cF_{0\o}(\kb+\qb,\kb_0;Q,\Q_s) -
\Theta(k-q) \cF_{0\o} (\kb,\kb_0;Q,\Q_s)
\right\}
\cr&
+ \delta^2(\kb-\kb_0)\,\Theta(Q-\Q_s)\,
\left[
1+\frac{\bas}{\o}\,\ln \frac{k_0^2}{\Q_s^2}\,\Theta(k_0-\Q_s)
\right]
\,.
}
\eeq
The kernel of this equation is regular for the collinear limit
$q \to 0$ so that in the integral we do not need the cutoff $\Q_s$.
The cutoff can be removed from the inhomogeneous term by defining
the unintegrated structure function
for which we can take the limit $\Q_s/Q \to 0$
\beq\label{BFKLreg}
\tilde \cF_{0\o}(\kb,\kb_0) \equiv
\cF_{0\o}(\kb,\kb_0;Q,\Q_s)
\left[
1 + \frac {\bas}{\o}\ln\frac{\kb_0^2}{\Q_s^2}\,\Theta(k_0-\Q_s)
\right]^{-1}
\,,
\;\;\;\;\;\;\;
\frac{\Q_s}{Q} \to 0
\,.
\eeq
This distribution satisfies the BFKL equation in \re{BFKL}
with the inhomogeneous term $\delta^2(\kb-\kb_0)$.

We want to show that, due to the cancellation of collinear
singularities, the unintegrated distribution
$\cA_\o(\kb,\kb_0;\bq;Q,\Q_s)$ obtained in the present formulation,
satisfies to the present accuracy the BFKL equation.  From \re{calA0}
we obtain (see \cite{C,CFM})
\beq\label{inteq01}
\eqalign{
&
\cA_\o(\kb,\kb_0;\bq;Q,\Q_s)=
\delta^2(\kb-\kb_0)\Theta(\bq-\Q_s)\,
\cr&
+ \int_{0}^{Q^2}
\;\frac{d^2q}{\pi q^2} \;dz \,z^\o \frac {\bas}{z} \;\Delta(z,q,k) \;
\Theta(\bq -zq) \;\cA_\o(\kb+\qb,\kb_0;q;Q,\Q_s)\,, }
\eeq
Integrating by part
$$
\int_0^1 dz \,z^\o\,\frac{\bas}{z}\, \Delta(z,q,k)\,
\Theta(\bq-zq) = \frac{\bas}{\o} \left\{ \Theta(\bq-q) - \int_0^1 dz
\,z^\o \;\frac{\partial}{\partial z} \left(
\Delta(z,q,k)\,\Theta(\bq-zq) \right) \right\}
$$
we obtain (the $Q$ and $\Q_s$ dependence is understood)
\beq\label{inteq02}
\eqalign{
\cA_\o(\kb,\kb_o;\bq)
&
= \frac{\bas}{\o} \int_{\Q_s}^{Q^2}
\;\frac{d^2q}{\pi q^2} \left\{ \cA_\o(\kb+\qb,\kb_0;q) -\Theta(k-q)\,
\cA_\o(\kb,\kb_0;q') \right\} \;+\; \delta_\o(\kb,\bq)
\cr&
+
\delta^2(\kb-\kb_0)\,\Theta(\bq-\Q_s)\,
\left[
1+\frac{\bas}{\o}\,\ln \frac{k_0^2}{\Q_s^2}\,\Theta(k_0-\Q_s)
\right]
\,, }
\eeq
with $q' \equiv$ min $(q,\bq)$ and
$$
\delta_\o(\kb,\bq) \equiv \frac{\bas}{\o}
\int_{\bq^2}^{Q^2} \;\frac{d^2q}{\pi q^2} \cA_\o(\kb+\qb,\kb_o;q) \,
\left[ \left(\frac {\bq}{q} \right)^\o\Delta(\frac{\bq}{q},q,\kb)-1
\right] \,.
$$
As before we remove the $\Q_s$ dependence in the
inhomogeneous term by defining the distribution
\beq\label{cAreg}
\tilde
\cA_{\o}(\kb,\kb_0;\bq;Q,\Q_s) \equiv \cA_{\o}(\kb,\kb_0;\bq;Q,\Q_s)
\left[ 1+\frac{\bas}{\o}\,\ln \frac{k_0^2}{\Q_s^2}\,\Theta(k_0-\Q_s)
\right]^{-1} \,.
\eeq
In the equation for this distribution we can remove the
cutoff $\Q_s$ and take the limit $\Q_s/Q \to 0$.
The contribution $ \delta_\o(\kb,\bq)$ vanishes for $\bq \to Q$
and does not contains leading contributions
given by the powers $\as^n/\o^n$.  By neglecting $ \delta_\o(\kb,\bq)$
we have that the integral equation for $\tilde \cA$ becomes
the BFKL equation and in these limits we can
approximate
\beq\label{cAreg1}
\tilde \cA_{\o}(\kb,\kb_0;\bq;Q,\Q_s) \simeq
\tilde \cF_{\o}(\kb,\kb_0)
\,.
\eeq

\section{Associated parton distributions}
The coherent formulation can be used to compute not only the
structure function but also the associated distributions for
any value of $x$ large or small by generalizing the jet
calculus \cite{jetC}.
As discussed in the previous section the present formulation with
coherence is equivalent to the BFKL equation only for the fully
inclusive case ($\kb$-structure function), but different for
distributions of the associated radiation.

As illustration we consider the single inclusive parton
distribution with a given energy fraction $y$  in the hard process
of Fig.~1. The relevant distribution is described in Fig.~4 and is
given in terms of the distribution $\cA(x,\kb,\kb_0;Q;\bq,\Q_s)$
as follows (take $\kb_0=0$ for simplicity)
\beq\label{Sig}
\eqalign{
\Sigma(x,y,Q,\Q_s) =
&
\int dx'\, d^2\kb\, d^2\kb\,'
\frac {1}{x'z}\,
\cA\left( \frac{x}{x'z},\kb,\kb';\bq=Q;Q,zq \right)
\frac {1}{x'(1-z)}\,
D \left(\frac{y}{x'(1-z)},p_t\right)
\cr&\times
\int^{Q}_{\Q_s}
\frac{d^2q}{\pi q^2} \,dz \,\bas\, P(z,q,k')
\, \cA(x',\kb'+\bom p_t,\kb_0;q;Q,\Q_s)
\,,
}
\eeq
where $\pb_t=(1-z)\qb$ is the initial state emitted transverse
momentum,  $P(z,q,k)$ is the splitting function in \re{br} which
includes the non-Sudakov forma factor and
$D(y/x'',q)$ is the inclusive distribution of the final
state gluon of energy fraction $y$ emitted in a jet of energy fraction
$x''=(1-z)x'$ and the hard scale is $p_t=(1-z)q$ as required by
angular  ordering.
The maximum angular variable is set at $\bq=Q$ as required by \re{q}.
{}From this expression we can compute for instance the average
associated multiplicity.
Notice that in these cases one maximises the value of $p_t$.

Consider the region of small $x$ in which all $z$ variables are
small. Taking the moments one finds
\beq\label{Sig0}
\eqalign{
\Sigma_{\o,\o'}(Q,\Q_s) \equiv
&
\int dx \, x^\o\, dy \, y^{\o'} \; \Sigma(x,y;Q,\Q_s)
=
\int d^2\kb\, d^2\kb'\, \cA_{\o}(\kb,\kb';\bq=Q;Q,zq)
\; D_{\o'}(q)
\cr& \times
\int^{Q}_{\Q_s}
\frac{d^2q}{\pi q^2} dz \,z^\o(1-z)^{\o'}
\,\frac{\bas}{z} \Delta(z,q,k')
\cA_{\o+\o'}(\kb'+\qb,\kb_0;q;Q,\Q_s)
\,,
}
\eeq
where for small $x$ we have approximated $p_t \simeq q$.

It is easy to check that this distribution satisfies
the energy sum rule, namely one has
$$
\Sigma_{\o,1}(Q)=\int dx \,x^\o\, (1-x) \, F(x,Q)
=F_\o(Q)-F_{\o+1}(Q)\,.
$$
We take $\o'=1$ in \re{Sig0} and use the energy sum rule
$D_{\o'=1}(q)=1$. We have
\beq
\eqalign{
\Sigma_{\o,1}(Q) =
&
\int d^2\kb\, d^2\kb'
\, \cA_{\o}(\kb,\kb';\bq=Q;Q,zq)
\, \int^{Q^2}_{\Q_s^2}\frac{d^2q}{\pi q^2}
\, dz \,(z^\o-z^{\o+1})
\, \frac{\bas}{z} \, \Delta(z,q,\kb')
\cr&\times
\, \cA_{\o+1}(\kb'+\qb,\kb_0;q;Q,\Q_s)
}
\,.
\eeq
For the second term we use the small $x$ evolution equation
\re{eveq0}. Changing integration variable from $q$ to $\bq=zq$ we have
$$
\bq^2 \frac{\partial}{\partial \bq^2}
\cA_{\o+1} (\kb',\kb_0;\bq;Q,\Q_s)=
\int_0^1 dz
\, z^{\o+1}
\, \frac {\bas}{z} \, \Delta(z,{\bq}/{z},\kb')
\; \cA_{\o+1}(\kb'+\bar \qb/z,\kb_0;\bq/ z;Q,\Q_s)
\, \Theta(Q-{\bq}/{z}) \,.
$$
For the first term we use the evolution equation in the minimum
angle
$$
q^2 \frac{\partial}{\partial q^2}\cA_\o(\kb,\kb'+\qb;\bq;Q,q)=
-\int_0^1 dz
\, z^\o
\, \cA_\o(\kb,\kb';\bq;Q,zq)
\, \frac {\bas}{z} \, \Delta(z,q,\kb')
\Theta(Q-q) \,.
$$
Finally, changing variables $\kb'+\qb$ to $\kb'$ in this second
contribution we have
$$
\Sigma_{\o,1}(Q)=
-\int d^2 \kb\, d^2 \kb'
\int_{Q^2_s}^{Q^2} dq^2 \frac{\partial }{\partial q^2}
\left\{
\cA_\o(\kb,\kb';\bq=Q;Q,q)\;
\cA_{\o+1}(\kb',\kb_0;q;Q,\Q_s)
\right\}
\,.
$$
By using the boundary condition
$$
\cA_\o(\kb,\kb_0;\bq=\Q_s;Q,\Q_s)=\delta^2 (\kb-\kb_0)\,,
$$
and
$$
F_\o(Q)=\int d^2 \kb \,\cA_\o(\kb,\kb_0;\bq=Q;Q,\Q_s)\,,
$$
we find that the sum rule is satisfied.

It is interesting to compare \re{Sig0} with the corresponding
expression obtained neglecting coherence by extending the
multi-Regge phase space \cite{Bart,Pe}.
As an example consider the case in which we neglect the final
state radiation, \ie we set $D_{\o'}(q)=1$ in \re{Sig0}. If we take
$\o'=0$ we are computing the average number of emitted initial sate
gluons. This average has been computed perturbatively up to the
fourth loop in the previous Section. This show that the high singular
terms $\as^n/\o^{2n-m}$  with $m<n$, present in the formulation with
coherence, do not cancel. Therefore, in this case the approximation of
substituting angular ordering with multi-regge region is not valid.

In general we can identify in \re{Sig0} the origin of these high
singular terms.
{}From \re{cAreg} and \re{cAreg1} we have that
the collinear dependence on the cutoff $zq$
in the first distribution $\cA$ inside the integral of \re{Sig0}
is given by a factor
$$
1+\frac{2\bas}{\o}\ln\frac{k'}{zq}\,\Theta(k'-zq)
\,.
$$
This factor is not compensated by the integration over the non-Sudakov
form factor due to angular ordering constraints $k>zq$ and this
generates $\as^n/\o^{2n-m}$ with $m<n$ contributions not present in
the multi-Regge region.
The conclusion is that for the associated distributions in general
angular ordering can not be approximated by multi-Regge regions.

\subsection{Jet calculus at small $x$}

The expression of the single inclusive distribution in \re{Sig0} can be
generalized by using the jet calculus algorithm. This algorithm is
based on the fact that the branching process is factorized.
For instance the generating functional of $y$ distributions can be
obtained as follow.
Consider first the generating functional $G_t[1,q,u]$ of inclusive
distribution in a gluon jet of energy fraction $1$ and angular
variable $q$ which depends on the function $u(y)$.
This functional is normalized by $ G_t[1,q,1]=1$ and
the single inclusive $y$ distribution is given by
$$
D(y,Q)=\frac{\delta}{\delta u(y)} \; G_t[1,q,u]\,|_{u=1}\,.
$$
The general $n$-gluon inclusive distributions are obtained by
further functional differentiation at $u(y)=1$.
Then we define the corresponding functional for the hard process
at scale $Q$ of initial state radiation at small $x$.
We define the space like functional
$G_s[x,\kb,\kb_0;\bq;Q,\Q_s;u]$
with the previous meaning for the variables.
By using the factorization of the branching process this functional
satisfies the following evolution equation
\beq\label{Jcalc}
\eqalign{
&
G_s[x,\kb,\kb_0;\bq;Q,\Q_s;u]=
\delta(1-x)\delta^2(\kb-\kb_0)\Theta(\bq-\Q_s)+
\int^{Q^2}_{0}\,\Theta(\bq-zq)\, \frac{d^2q}{\pi q^2}\frac{dz}{z}
\cr& \times
\left\{
\frac{\bas}{z}\Delta(z,q,\kb)\;G_t[x',q,u_{x'}]
\right\}
\;
G_s[\frac x z ,\kb+\qb,\kb_0;q;Q,\Q_s,u]
\,,
}
\eeq
where $x'=(1-z) x/z$ and $u_x(y)=u(xy)$.
{}From \re{inteq0} we have that at $u(y)=1$ this functional is the unintegrated
distribution $\cA(x,\kb,\kb_0;\bq,\Q_s)$.
The single inclusive distribution in \re{Sig} is obtained by
differentiation
$$
\Sigma(x,y,Q) = \int d^2 \kb d^2 \kb'\;
\frac{\delta}{\delta u(y)} \; G_s[x,\kb,\kb_o;Q;Q,\Q_s,u]\,|_{u=1}\,.
$$
To see that this distribution coincides with the one in \re{Sig0}
one proceeds in some formal way exploiting factorization
which is the bases of jet calculus.
We write \re{Jcalc} in the formal way
$$
G_s[u]=1+K[u]\cdot G_s[u]\,,
$$
where the kernel $K[u]$ corresponds to the distribution in the curly
bracket in \re{Jcalc}. Taking the functional derivative one has
$$
\Sigma= \frac{\delta}{\delta u(y)} \; G_s[u]\,|_{u=1}=
G_s[u]\cdot \frac{\delta K[u]}{\delta u(y)}\cdot G_s[u]\,|_{u=1}
\,.
$$
The expression in \re{Sig} is obtained from the fact that $G_s[1]$ is
into account that $G_s[1]$ is the distribution $\cA$.

\section{Conclusions}

We recall the main features of the analysis here presented for small
and large $x$.

1) At the level of the fully inclusive distributions at small $x$
($\kb$-structure function) the dominant region of phase space is the
multi-Regge region with strong $x_i$-ordering and with transverse
momenta of the same order. This is due to the complete cancellation
of the collinear singularities which arises when two emitted
transverse momenta are very different.
This property is the basis of the BFKL equation and of the fact that
the hard scale associated to collinear singularities is lost.

2) For associated distributions the collinear singularities do not
cancel in general and the dominant region of phase space is
determined by angular ordering.
The analysis of the associated distributions for small $x$
requires the complete structure of collinear singular terms, \ie
no collinear approximations.
This level of accuracy can be actually attained for \xt0 by using
soft gluon factorization.
While for the amplitudes the soft gluon factorization does not require
any collinear approximation, for the distributions one is usually forced
to introduce collinear approximation in order to perform the
multi-gluon colour algebra.
For \xt0 one avoids this approximation and the basis of this fact is
in eqs.~\re{ne} and \re{noncoll}.
Due to the real-virtual cancellation of soft singularities and
to the coherent structure of the eikonal current, one finds that
the distribution of a soft gluon emitted by a jet of partons
does not vanish for \xt0 only if the jet of partons are confined into
an angular cone with aperture of order $x$.
Therefore for \xt0 we can use collinear estimates.
Notice that the approximation involved is the soft approximation,
not the collinear one.

3) QCD coherence has the same origin for small and large $x$ and
and in both regions the emission factorizes.
Therefore one can write a unified evolution equation which resums
all IR singularities which at small $x$ does not involve any
collinear approximation while at large $x$ is accurate only to
leading collinear level.
Contributions from non soft radiation as well as contributions
from quarks can be included in a natural way (see also \cite{Pe}).
In this equation, the most important difference between the two
regions of small and large $x$ is the presence of the non-Sudakov
or hard Regge form factor \re{nsud} for \xt0.
For finite $x$ this form factor gives a finite correction while
for \xt0 this form factor cancels all the collinear singularities
in the $\kb$-structure function. Thus at this fully inclusive level
angular ordering becomes equivalent to the multi-Regge phase space.
For associated distribution instead the cancellation does not take
place and one can not neglect angular ordering. Thus from these
distributions one can reveal the coherence of QCD radiation in
small $x$ processes.
To study these associated distributions we have formulated a
unified branching algorithm for DIS processes which allows one to
compute associated distributions in all regions of $x$.

\vskip .3 true cm
\noindent{\bf Acknowledgements}\vskip .1 true cm
I am most grateful for valuable discussions with
J.\ Bartels,
S.\ Catani,
M.\ Ciafaloni,
Yu.L.\ Dokshitzer,
I.\ Kwieci\'nski,
A.H.\ Mueller,
and
B.R.\ Webber.

{\bf Appendix A}

Take $n=m+\ell$ and consider the case in which the $\ell$ softest
gluons $q_{m+1},\cdots,q_{m+\ell}$ are virtual while $q_{m}$ is real
\beq\label{a1}
y_{m+\ell}\ll \cdots y_{m+1}\ll y_{m}
\,.
\eeq
Since the quantities $Y_{R_n}$ and $\kb_{R_n}$ do not involve
the $\ell$ softest momenta the two corresponding $\delta-$functions
can be factorized and one can sum over these softest momenta and
reconstruct the form factors.
{}From \re{rr} we have in the region \re{a1}
\beq\label{a2} \eqalign{
&
g_s^{2\ell}\,\prod_1^\ell (dq_{m+\ell})
f_{R_n}(q_1,\cdots,q_{m+\ell})\,
\cr&
=
\prod_1^\ell
\left\{
-\bas\,
\frac{dy_{m+i}}{y_{m+i}}\frac{d^2q_{m+i}}{\pi q_{m+i}}
\,[1-\Theta(y_{m+i}-x)\,\Theta(q_{m+i}-k)]
\right\}
\,f_{R_m}(q_1,\cdots,q_{m})
\,,
}
\eeq
Integrating over the region \re{a1} and summing over $\ell$ one finds
\beq\label{a3}
\eqalign{
&
\int \Theta^Y_{1\cdots n}
\, \sum_{\ell=0}^\infty
\, g_s^{2\ell}\,\,\prod_1^\ell (dq_{m+i})
\,f_{R_n}(q_1,\cdots,q_{n})
\cr&
=
\Theta^Y_{1\cdots m}
\, f_{R_m}(q_1,\cdots,q_{m})\, exp \left\{
-\bas \int^{y_m}\frac{dy}{y}\,\int^{Q^2}\frac{dp^2}{p^2}
+\bas \int^{y_m}_{x}\frac{dy}{y}\,\int^{Q^2}_{k^2}\frac{dp^2}{p^2}
\right\}
\,,
}
\eeq
where in \re{a1} we have $x_m= x+y_{m+\ell}+\cdots+y_{m+1} \sim
y_m$. Iterating this procedure to the next to soft emitted gluons we
obtain the result in \re{V}.

\vskip 1. true cm

{\bf Appendix B}

We consider the exchange of $y$ with $\xi$ ordering in \re{M2}.
The factor not obviously symmetric with respect to $y_i$ and $\xi_i$
is given by
\beq\label{b1}
\Theta^Y_{1 \cdots r} \prod_1^r T(y_i,y_{i+1},k_i)\,,
\;\;\;\;\;\;\;\;
T(y_i,y_{i-1},k_i)=
\exp\left\{
-\bas \int_{y_i}^{y_{i+1}} \frac {dy}{y}\int_{k^2}^{Q^2}
\frac{dp^2}{p^2}
\right\}
\,,
\eeq
where the $\Theta$-function corresponds the strong ordering region \re{yo}.
Here we have used $x_i \simeq y_{i+1}$, $x \equiv y_{n+1}$,
and $\kb_i=-\bom q_i+\kb_{i-1}$.
Since the quantity in \re{b1} is integrated over we can exchange
variable names as follows
\beq\label{b2}
\Theta^Y_{1 \cdots r} \prod_1^r T(y_i,y_{i+1},k_i)
\;\;
\Rightarrow
\;\;
\Theta^\xi_{r \cdots 1} \sum_{\mbox{perm.}}
\Theta^Y_{\ell_1 \cdots \ell_r}\prod_1^r
T(y_{\ell_{i}},y_{\ell_{i+1}},q_{\ell_1\cdots \ell_i})
\,,
\eeq
where
$\bom q_{\ell_1\cdots \ell_i} \equiv \bom
q_{\ell_1}+\cdots+ \bom q_{\ell_i}$.

Consider the case with two real gluons. There are two permutations
in \re{b2}. The first with $y_2 \ll y_1$ and $\xi_1<\xi_2$ gives
$y_1 \simeq 1$ and $x_1=x+y_2\simeq y_2$ so that
$$
\Theta^\xi_{21}
\Theta^Y_{12}\,T(y_1,y_2;q_1)\,T(y_2,x;q_{12})
\;\simeq\;
\Theta^\xi_{21}
\Theta^Y_{12}\,T(1,x_1;q_1)\,T(x_1,x;q_{12})
\,.
$$
The second permutation with the ordering $y_1 \ll y_2$
and $\xi_1<\xi_2$ gives $y_2\simeq 1$, $\;x_1=x+y_2\simeq 1$ and
$\;q_1\ll q_2$ so that we can write
$$
\Theta^\xi_{21}
\Theta^Y_{21}\,T(y_2,y_1;q_2)\,T(y_1,x;q_{12})
\;\simeq\;
\Theta^\xi_{21}
\Theta^Y_{21}\,T(1,x;q_{12})
\;\simeq\;
\Theta^\xi_{21}
\Theta^Y_{21}\,T(1,x_1;q_1)\,T(x_1,x;q_{12})
\,.
$$
The two contributions have the same form within the leading IR
accuracy and we conclude that we can substitute $y$-ordering
with $\xi$-ordering $$
\Theta^Y_{12}\,T(y_1,y_2;q_1)\,T(y_2,x;q_{12})
\;\;\Rightarrow\;\;
\Theta^\xi_{21}\, T(1,x_1;q_1)\,T(x_1,x;q_{12})
$$
Consider the case with three emitted gluons $r=3$.
After singling out, as in \re{b2}, the angular ordering
$\xi_1<\xi_2<\xi_3$ with $\Theta^\xi_{321}$ we have six permutations
of $y$ ordering configurations. Consider the contribution from the
fundamental permutation
$$
\Theta^\xi_{321}
\Theta^Y_{123}\,T(y_1,y_2;q_1)\,T(y_2,y_3;q_{12})\,T(y_3,x;q_{123})
\,.
$$
Since $\xi_1 < \xi_2 <\xi_3$ and $y_3\ll y_2\ll y_1$,
we have $y_1\simeq 1$, $\,x_2=x+y_3\simeq y_3$ and
$\, x_1=x_2+y_2\simeq y_2$ so that, factorizing
$\Theta^\xi_{321}\,\Theta^Y_{123}$, we have
$$
T(y_1,y_2;q_1)\,T(y_2,y_3;q_{12})\,T(y_3,x;q_{123})
\simeq
T(1,x_1;q_1)\,T(x_1,x_2;q_{12})\,T(x_2,x;q_{123})
\,.
$$
The same expression is obtained for all other permutations within
the leading IR accuracy. Consider for instance the contribution
$$ \Theta^\xi_{321}\,
\Theta^Y_{213}\,T(y_2,y_1;q_2)\,T(y_1,y_3;q_{12})\,T(y_3,x;q_{123})
\,,
$$
where $\xi_1 < \xi_2 <\xi_3$ and $y_3\ll y_1\ll y_2$.
We have $y_2\simeq 1$, $x_2=x+y_3\simeq y_3$,
$x_1=x_2+y_1\simeq 1$ and $ q_1 \ll q_2$.
After factorizing $\Theta^\xi_{321}\,\Theta^Y_{213}$, we have
$$
T(y_2,y_1;q_2)\,T(y_1,y_3;q_{12})\,T(y_3,x;q_{123})
\simeq
T(1,y_1;q_2)\,T(y_1,x_2;q_{12})\,T(x_2,x;q_{123})
$$
$$
\simeq
T(1,x_2;q_{12})\,T(x_2,x;q_{123})
\simeq
T(1,x_1;q_1)\,T(x_1,x_2;q_{12})\,T(x_2,x;q_{123})
\,.
$$
For the other permutations one proceeds as before and finds the same
expression so that one can substitute, to leading IR accuracy,
$$
\Theta^Y_{213}\,T(y_2,y_1;q_2)\,T(y_1,y_3;q_{12})\,T(y_3,x;q_{123})
\Rightarrow
\Theta^\xi_{321}\,T(1,x_1;q_1)\,T(x_1,x_2;q_{12})\,T(x_2,x;q_{123})
\,.
$$
The proof for the general case can be done by induction.
A general physical argument which allows the substitution of
$y-$ordering with $\xi-$ordering is given in \cite{CFM}.

\vskip 1. true cm

{\bf Appendix C}

The form in \re{V} of virtual corrections has been discussed
\cite{CFM} and will be recalled here for completeness.
{}From \re{V} we have
\beq
\ln V_r(p_1\cdots p_r)=
-\bas \int_{0^+}^1 \frac{dy}{y}\, \int_{0^+}^1 \frac{d\xi}{\xi}
+\sum_1^r \bas \int_{x_i}^{x_{i-1}} \frac{dy}{y}\,
\int_{k_i^2}^{Q^2} \frac{d p^2}{p^2}
\,,
\eeq
where in the first term we introduced the angular
variable $\xi$ of virtual gluon.
This first term with the singularity for $ y, \xi \to 0$ can be
expressed as sum of Sudakov form factors which regularize the
singularities in the emitted phase space for $y_i, q_i \to 0$.
To this end we introduce in the virtual correction the same IR cutoff
as in the real emission, namely for the virtual transverse momenta we
set the cutoff $Q_0$.
In the angular ordered region of the emitted phase space
$$
\xi_0<\xi_1<\cdots<\xi_r<1
\,,
$$
we have $q_i=x_{i-1}\sqrt{s\xi_i}$ and introduce the virtual momentum
$p=y\sqrt{s\xi}>Q_0$.
We have
\beq
\ln V_r(p_1\cdots p_r)=
-\sum_0^r \bas
\int_{0^+}^{x_i} \frac{dy}{y}\,
\int_{\xi_i}^{\xi_{i+1}} \frac{d\xi}{\xi}\,\Theta(y\sqrt{s\xi}-q_0)
+\sum_1^r \bas \int_{x_i}^{x_{i-1}} \frac{dy}{y}\,
\int^{k_i^2} \frac{d p^2}{p^2}\
\, \Theta(p-y\sqrt{s\xi})
\,.
\eeq
Introducing in the first sum the variables
$y=(1-z)x_i$ and $q=x_i\sqrt{s\xi}$
and in the second the variable $y=zx_{i-1}$,
we deduce the final result in \re{V11}.
\par 

\eject

\newpage
\begin{figcap}

\item
Deep inelastic scattering at parton level. The dotted line represents
the off shell photon $q$.
For the dominat small $x$ contribution the incoming $p$ and the
outgoing $p_i$ partons are gluons and the recoiling system is a
quark-antiquark pair of momentum $p'$.

\item
Kinematical diagram for parton emission.

\item
Graphical representation of the elementary vertex factorization.

\item
Jet calculus diagram for the associated single inclusive distribution
of parton $q$ with transverse momentum $\bom p_t$ and energy fraction
$y$. The exchanged partons $k$, $k'$ and $k'+q$  have
transverse momenta and energy fractions given by
$\kb, \, x\,$, $\kb', \, zx'\,$ and $\kb'+\bom p_t, \, x'\,$
respectively.
\end{figcap}

\newpage
\newcount\x \newcount\y \newcount\p \newcount\q
\newcount\n \newcount\l \newcount\m
\setlength{\topmargin}{-1.5 cm}
\setlength{\evensidemargin}{.0 cm}
\setlength{\oddsidemargin}{.0 cm}
\setlength{\textheight}{9.5 in}
\setlength{\textwidth}{6.5 in}
\parskip = 2ex
\newskip\humongous \humongous=0pt plus 1000pt minus 1000pt
\def\caja{\mathsurround=0pt} \def\eqalign#1{\,\vcenter{\openup1\jot
\caja	\ialign{\strut \hfil$\displaystyle{##}$&$
\displaystyle{{}##}$\hfil\crcr#1\crcr}}\,} \newif\ifdtup
\def\panorama{\global\dtuptrue \openup1\jot \caja
\everycr{\noalign{\ifdtup \global\dtupfalse	\vskip-\lineskiplimit
\vskip\normallineskiplimit	\else \penalty\interdisplaylinepenalty \fi}}}
\def\eqalignno#1{\panorama \tabskip=\humongous
\halignto\displaywidth{\hfil$\displaystyle{##}$
\tabskip=0pt&$\displaystyle{{}##}$\hfil
\tabskip=\humongous&\llap{$##$}\tabskip=0pt	\crcr#1\crcr}}

\def\bom#1{\mbox{\bf{#1}}} \def\half{\mbox{\small $\frac{1}{2}$}}
\def\tird{\mbox{\small $\frac{1}{3}$}}
\def\ltap{\raisebox{-.4ex}{\rlap{$\sim$}} \raisebox{.4ex}{$<$}}
\def\gtap{\raisebox{-.4ex}{\rlap{$\sim$}} \raisebox{.4ex}{$>$}}
\def\VEV#1{\left\langle #1\right\rangle}
\def\frac#1#2{ {{#1} \over {#2} }}
\def\etal{\hbox{\rm et al.}}
\def\as{\alpha_S}
\def\asb{\bar \alpha_S}
\def\blank#1{{\hbox {\hskip #1 true cm}}}   
\def\abs#1{\left| \: #1 \: \right|}%


\def\dot#1#2#3#4#5{
\multiput(#1,#2)(#3,#4){#5}{.}}


\def\dashdvec#1#2#3#4#5{          
\n=#5 \divide \n  by 6  \x = #3
\m = \n \divide \m by 2 \multiply \m by 2
\advance \m by -\n \multiply \m by 3
\multiply \x by 6 \y = #4 \multiply \y by 6
\p = #5 \divide \p by 2 \advance \p by \m  \q = \p
\multiply \p by #3 \multiply \q by #4
\advance \p by #1 \advance \q by #2
\multiput(#1,#2)(\x,\y){\n}{\line(#3,#4){4}}
\put (\p,\q){\vector (#3,#4) {3}}}


\def\live#1#2#3#4#5{
\n = #5 \divide \n  by 2  \advance \n by 1
\put(#1,#2){\line (#3,#4){#5}}
\put(#1,#2){\vector (#3,#4){\n}}
}


\def\boxc#1#2#3#4#5{
\put (#1,#2){\makebox(#3,#4){\parbox{#3 \unitlength}{\centering #5}}}}


\def\boxoval#1#2#3#4#5{
  \x = #3 \multiply \x by 2  \y = #4 \multiply \y by 2
    { \put (#1,#2) {\oval (\x,\y)}
      \l = #1  \advance \l by -#3   \m = #2 \advance \m by -#4
      \p = -#3  \divide \p by 3  \advance \p by \x
      \put (\l,\m){\makebox(\x,\y){\parbox{\p \unitlength}{\centering #5}}}
                                   }}
\def\figone{
\setlength {\unitlength}{.9 mm}
\begin{center}

\begin{picture}(75,100)
\thicklines
\dashdvec{5}{9}{1}{1}{18}
\put (25,29)  {\circle{11.5}}
\put (25,34.5){\vector (0,1){10}}
\put (50.5,29){\vector (-1,0){10}}
\put (29,33)  {\vector (1,1){7}}
\put (27,34.5){\vector (1,2){5}}
\put (32,44.5){\line (1,2){4}}
\put (25,44)  {\line (0,1){10}}
\put (30.5,29){\line (1,0){25}}
\put (35,39)  {\line (1,1){7}}
\put(1,9)   {$q$}
\put (1,50) {$ p' $}
\put (26,54){$p_r$}
\multiput (27,50)(2,- .5){3}{$.$}
\put (38,52){$ p_2 $}
\put (44,46){$p_1$}
\put (53,23){$p$}
\put (21.1,33.1){\vector (-1,1){8}}
\put (21,33)    {\line (-1,1){16}}
\put (21.2,33.2){\line (-1,1){16}}
\boxc{0}{0}{56}{6}{\sl Figure~1}
\end{picture}
\end{center}
}
\def\figtwo{
\begin{center}
\begin{picture}(101,100)
\put(0,8){
\begin{picture}(80,50)
\thicklines
\live {15}{12}{-1}{0}{15}
\put (5,6) {$k_r$}
\live {35}{12}{-1}{0}{20}
\put (27,6) {$k_2$}
\live {50}{12}{-1}{0}{15}
\put (45,6) {$k_1$}
\live {65}{12}{-1}{0}{15}
\put (55,6) {$p$}
\live {15}{12}{0}{1}{15}
\put (15,30) {$p_r$}
\live {35}{12}{0}{1}{15}
\put (35,30) {$p_2$}
\live {50}{12}{0}{1}{15}
\put (50,30) {$p_1$}
\end{picture}}
\dot{22}{30}{3.5}{0}{3}
\boxc{0}{0}{50}{6}{\sl Figure~2}
\end{picture}
\end{center}}
\def\figthree{
\setlength {\unitlength}{1 mm}
\begin{center}
\begin{picture}(101,100)
\put(10,5){
\begin{picture}(25,50)
\thicklines
\put (11,27) {\circle*{4}}
\live {24}{27}{-1}{0}{12}
\put(4,14)  {\vector (1,2) {2.6}}
\put(7.5,21){\line (1,2) {3}}
\put (0,11) {$q$}
\put (15,21) {$ k \; \mu$}
\put(10,27){\line (-1,2){6}}
\put(10.2,27.3){\line (-1,2){6}}
\live {10.1}{27.2}{-1}{2}{6}
\put (0,41) {$p'$}
\end{picture}}
\boxc{13}{10}{20}{8}{$ V^\mu (q,k)$}
\boxc{0}{0}{101}{6}{\sl Figure~3}
\put(53,5){
\begin{picture}(39,50)
\thicklines
\live {3}{27}{1}{0}{11}
\put (2,21) {$k \; \mu$}
\live {39}{27}{-1}{0}{11}
\put (36,21) {$p$}
\boxoval{21}{27}{7}{5}{}
\live {21}{32}{0}{1}{14}
\put (22,47) {$p_2$}
\live {17}{32}{-1}{2}{7}
\put (11.5,47) {$p_r$}
\live {25}{32}{1}{2}{7}
\put (33.5,47) {$p_1$}
\dot{23}{44}{2}{- .4}{3}
\end{picture}}
\boxc{50}{10}{45}{8}{$ {\bf m}^{(r)}_\mu (p,k; p_1 \ldots p_r)$}
\end{picture}
\end{center}}
\def\figfour{
\begin{center}
\begin{picture}(101,100)
\thicklines
\put(10,10){
\begin{picture}(90,35)
\thicklines
\live {43}{12}{0}{1}{10}
\put   (43,26)  {\circle{8}}
\live {43}{30}{0}{1}{5}
\put   (43,38) {$q$}
\live {14}{12}{-1}{0}{15}
\put (6,6) {$k$}
\live {43}{12}{-1}{0}{15}
\put (34,6) {$k'$}
\live {58}{12}{-1}{0}{15}
\put (46,6) {$k'+q$}
\live {82}{12}{-1}{0}{10}
\put (77,6) {$p$}
\boxoval{21}{12}{7}{5}{}
\live {21}{17}{0}{1}{14}
\live {17}{17}{-1}{2}{7}
\live {25}{17}{1}{2}{7}
\boxoval{65}{12}{7}{5}{}
\live {65}{17}{0}{1}{14}
\live {61}{17}{-1}{2}{7}
\live {69}{17}{1}{2}{7}
\end{picture}}
\boxc{0}{0}{101}{6}{\sl Figure~4}
\end{picture}
\end{center}}
\figone
\figtwo
\figthree
\figfour
\end{document}
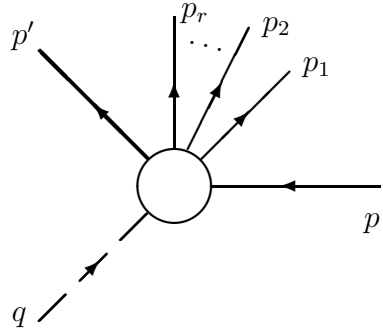
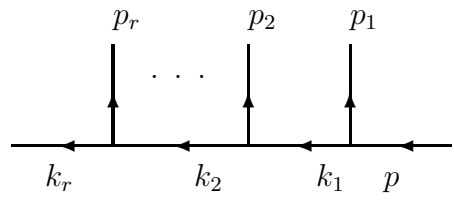
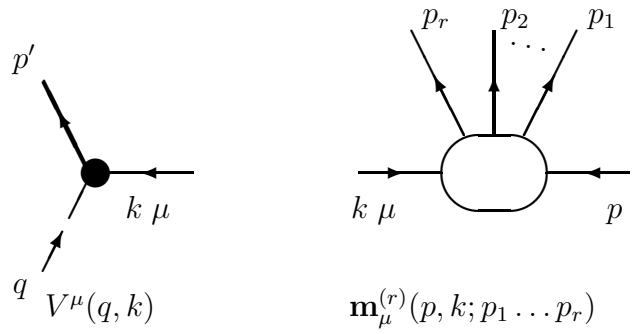
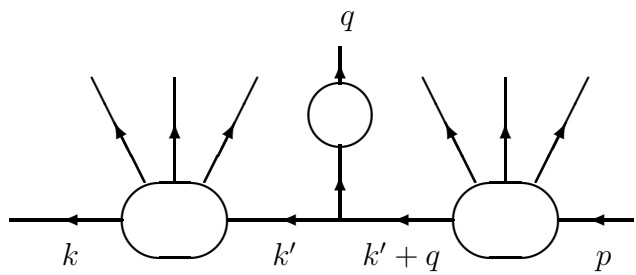
\end{document}